\documentclass[nofootinbib,groupedaddress,superscriptaddress,unsortedaddress,showpacs]{revtex4-1}
\usepackage{amsmath,amsfonts,slashed}
\usepackage{epsfig,epsf}
\usepackage{hyperref}
\usepackage{float}
\restylefloat{table}

\usepackage[dvips]{color}
\usepackage[normalem]{ulem}
\definecolor{darkgreen}{cmyk}{1,0,1,0.4}

\definecolor{darkred}{cmyk}{0,1,1,0.4}

\usepackage{slashed}

\newcommand{\newc}{\newcommand}
\def\preprint{{preprint}}
\def\Ord{\lower .7ex\hbox{$\;\stackrel{\textstyle <}{\sim}\;$}}
\def\OOrd{\lower .7ex\hbox{$\;\stackrel{\textstyle >}{\sim}\;$}}

\newc{\order}{{\cal O}}

\newc{\be}{\begin{equation}}
\newc{\ee}{\end{equation}}
\newc{\br}{\begin{eqnarray}}
\newc{\er}{\end{eqnarray}}
\newc{\ba}{\begin{array}}
\newc{\ea}{\end{array}}
\newc{\bi}{\begin{itemize}}
\newc{\ei}{\end{itemize}}
\newc{\bn}{\begin{enumerate}}
\newc{\en}{\end{enumerate}}
\newc{\bc}{\begin{center}}
\newc{\ec}{\end{center}}
\newc{\ul}{\underline}
\newc{\ra}{\rightarrow}
\newc{\lra}{\longrightarrow}
\newc{\wt}{\widetilde}
\newc{\til}{\tilde}

\newc{\wh}{\widehat}
\newc{\ti}{\times}
\newc{\Dir}{\kern -6.4pt\Big{/}}
\newc{\Dirin}{\kern -10.4pt\Big{/}\kern 4.4pt}
\newc{\DDir}{\kern -10.6pt\Big{/}}
\newc{\DGir}{\kern -6.0pt\Big{/}}
\newc{\sig}{\sigma}
\newc{\sigmalstop}{\sig_{\lstoppair}}
\newc{\Sig}{\Sigma}  
\newc{\del}{\delta}
\newc{\Del}{\Delta}
\newc{\lam}{\lambda}
\newc{\Lam}{\Lambda}
\newc{\gam}{\gamma}
\newc{\Gam}{\Gamma}
\newc{\eps}{\epsilon}
\newc{\Eps}{\Epsilon}
\newc{\kap}{\kappa}
\newc{\Kap}{\Kappa}
\newc{\modulus}[1]{\left| #1 \right|}
\newc{\eq}[1]{(\ref{eq:#1})}
\newc{\eqs}[2]{(\ref{eq:#1},\ref{eq:#2})}
\newc{\etal}{{\it et al.}\ }
\newc{\ibid}{{\it ibid}.}
\newc{\ibidem}{{\it ibidem}.}
\newc{\eg}{{\it e.g.}\ }
\newc{\ie}{{\it i.e.}\ }

\newc{\nonum}{\nonumber}
\newc{\lab}[1]{\label{eq:#1}}
\newc{\dpr}[2]{({#1}\cdot{#2})}
\newc{\lt}{\stackrel{<}}
\newc{\gt}{\stackrel{>}}
\newc{\lsimeq}{\stackrel{<}{\sim}}
\newc{\gsimeq}{\stackrel{>}{\sim}}
\def\lsim{\buildrel{\scriptscriptstyle <}\over{\scriptscriptstyle\sim}}
\def\gsim{\buildrel{\scriptscriptstyle >}\over{\scriptscriptstyle\sim}}
\def\lapp{\mathrel{\rlap{\raise.5ex\hbox{$<$}}
                    {\lower.5ex\hbox{$\sim$}}}}
\def\gapp{\mathrel{\rlap{\raise.5ex\hbox{$>$}}
                    {\lower.5ex\hbox{$\sim$}}}}
\newc{\half}{\frac{1}{2}}

\newc{\bQ}{\ol{Q}}
\newc{\dota}{\dot{\alpha }}
\newc{\dotb}{\dot{\beta }}
\newc{\dotd}{\dot{\delta }}
\newc{\nindnt}{\noindent}

\newc{\matth}{\mathsurround=0pt}
\def\ML{\ifmmode{{\mathaccent"7E M}_L}
             \else{${\mathaccent"7E M}_L$}\fi}
\def\MR{\ifmmode{{\mathaccent"7E M}_R}
             \else{${\mathaccent"7E M}_R$}\fi}

\newc{\mr}{\mathrm}

\newc{\siminf}{\mbox{$_{\sim}$ {\small {\hspace{-1.em}{$<$}}}    }}
\newc{\simsup}{\mbox{$_{\sim}$ {\small {\hspace{-1.em}{$>$}}}    }}

\newc {\Zboson}{{\mathrm Z}^{0}}
\newc{\thetaw}{\theta_W}
\newc{\mbot}{{m_b}}
\newc{\mtop}{{m_t}}
\newc{\sm}{${\cal {SM}}$}
\newc{\as}{\alpha_s}
\newc{\aem}{\alpha_{em}}

\newc{\ppbar}{\mbox{$p\ol{p}$}}
\newc{\bbbar}{\mbox{$b\ol{b}$}}
\newc{\ccbar}{\mbox{$c\ol{c}$}}
\newc{\ttbar}{\mbox{$t\ol{t}$}}
\newc{\eebar}{\mbox{$e\ol{e}$}}
\newc{\zzero}{\mbox{$Z^0$}}

\newc{\wplus}{\mbox{$W^+$}}
\newc{\wminus}{\mbox{$W^-$}}
\newc{\ellp}{\ell^+}
\newc{\ellm}{\ell^-}
\newc{\elp}{\mbox{$e^+$}}
\newc{\elm}{\mbox{$e^-$}}
\newc{\elpm}{\mbox{$e^{\pm}$}}
\newc{\qbar}     {\mbox{$\ol{q}$}}

\newc{\Ebar}{{\bar E}}
\newc{\Dbar}{{\bar D}}
\newc{\Ubar}{{\bar U}}
\newc{\susy}{{{SUSY}}}
\newc{\msusy}{{{M_{SUSY}}}}

\def\photino{\ifmmode{\mathaccent"7E \gam}\else{$\mathaccent"7E \gam$}\fi}
\def\taugluino{\ifmmode{\tau_{\mathaccent"7E g}}
             \else{$\tau_{\mathaccent"7E g}$}\fi}
\def\mphotino{\ifmmode{m_{\mathaccent"7E \gam}}
             \else{$m_{\mathaccent"7E \gam}$}\fi}
\newc{\gl}   {\mbox{$\wt{g}$}}
\newc{\mgl}  {\mbox{$m_{\gl}$}}

\def \chonep {{\wt\chi_1^+}}

\def \ch2p {{\wt\chi_2^+}}
\def \chonem {{\wt\chi_1^-}}
\def \ch2m {{\wt\chi_2^-}}

\def \chonjpm{{\wt\chi_j}^{\pm}}
\def \chonepm{{\wt\chi_1}^{\pm}}

\def \mchonepm{m_{\chonepm}}

\def \chtwopm{{\wt\chi_2}^{\pm}}
\def \mchtwopm{m_{\chtwopm}}
\newc{\dmchi}{\Delta m_{\wt\chi}}

\def \lspi{\wt\chi_i^0}

\def \lspone{\wt\chi_1^0}
\def \mlspone{m_{\lspone}}
\def \lsptwo{\wt\chi_2^0}
\def \mlsptwo{m_{\lsptwo}}
\def \lspthree{\wt\chi_3^0}
\def \mlspthree{m_{\lspthree}}
\def \lspfour{\wt\chi_4^0}
\def \mlspfour{m_{\lspfour}}

\newc{\sele}{\wt{\mathrm e}}
\newc{\sell}{\wt{\ell}}

\def \snu{\wt{\nu}}

\newc{\snue}     {\mbox{$ \wt{\nu_e}$}}
\newc{\smu}{\wt{\mu}}
\newc{\stau}{\wt{\tau}}
\newc {\nuL} {\wt{\nu}_L}
\newc {\nuR} {\wt{\nu}_R}
\newc {\snub} {\bar{\wt{\nu}}}
\newc {\eL} {\wt{e}_L}
\newc {\eR} {\wt{e}_R}
\def \slepl{\wt{l}_L}
\def \mslepl{m_{\slepl}}
\def \slepr{\wt{l}_R}
\def \mslepr{m_{\slepr}}
\def \stau{\wt\tau}

\def \sq{\wt{q}}

\newc{\msqot}  {\mbox{$m_(\sq_{1,2} )$}}
\newc{\sqbar}    {\mbox{$\bar{\wt{q}}$}}
\newc{\ssb}      {\mbox{$\squark\ol{\squark}$}}
\newc {\qL} {\wt{q}_L}
\newc {\qR} {\wt{q}_R}
\newc {\uL} {\wt{u}_L}
\newc {\uR} {\wt{u}_R}
\def \ul{\wt{u}_L}

\newc {\dL} {\wt{d}_L}
\newc {\dR} {\wt{d}_R}
\newc {\cL} {\wt{c}_L}
\newc {\cR} {\wt{c}_R}
\newc {\sL} {\wt{s}_L}
\newc {\sR} {\wt{s}_R}
\newc {\tL} {\wt{t}_L}
\newc {\tR} {\wt{t}_R}
\newc {\stb} {\ol{\wt{t}}_1}
\newc {\sbot} {\wt{b}_1}
\newc {\msbot} {m_{\sbot}}
\newc {\sbotb} {\ol{\wt{b}}_1}
\newc {\bL} {\wt{b}_L}
\newc {\bR} {\wt{b}_R}

\newc{\csquark}  {\mbox{$\wt{c}$}}
\newc{\csquarkl} {\mbox{$\wt{c}_L$}}
\newc{\mcsl}     {\mbox{$m(\csquarkl)$}}

\newc {\stopl}         {\wt{\mathrm{t}}_{\mathrm L}}
\newc {\stopr}         {\wt{\mathrm{t}}_{\mathrm R}}
\newc {\stoppair}      {\wt{\mathrm{t}}_{1}
\bar{\wt{\mathrm{t}}}_{1}}
\def \lstop{\wt{t}_{1}}

\def \lstoppair{\lstop\lstop^*}

\newc{\tsquark}  {\mbox{$\wt{t}$}}
\newc{\ttb}      {\mbox{$\tsquark\ol{\tsquark}$}}
\newc{\ttbone}   {\mbox{$\tsquark_1\ol{\tsquark}_1$}}

\newc{\mix}{\theta_{\wt t}}
\newc{\cost}{\cos{\theta_{\wt t}}}
\newc{\sint}{\sin{\theta_{\wt t}}}
\newc{\costloop}{\cos{\theta_{\wt t_{loop}}}}

\newc{\mixsbot}{\theta_{\wt b}}

\newc{\tb}{\tan\beta}
\newc{\cb}{\cot\beta}
\newc{\vev}[1]{{\left\langle #1\right\rangle}}

\newc{\mhalf}{m_{1/2}}
\newc{\mzero} {\mbox{$m_0$}}
\newc{\azero} {\mbox{$A_0$}}

\newc{\lb}{\lam}
\newc{\lbp}{\lam^{\prime}}
\newc{\lbpp}{\lam^{\prime\prime}}
\newc{\rpv}{{\not \!\! R_p}}
\newc{\rpvm}{{\not  R_p}}
\newc{\rp}{R_{p}}
\newc{\rpmssm}{{RPC MSSM}}
\newc{\rpvmssm}{{RPV MSSM}}

\newc{\sbyb}{S/$\sqrt B$}
\newc{\pelp}{\mbox{$e^+$}}
\newc{\pelm}{\mbox{$e^-$}}
\newc{\pelpm}{\mbox{$e^{\pm}$}}
\newc{\epem}{\mbox{$e^+e^-$}}
\newc{\lplm}{\mbox{$\ell^+\ell^-$}}

\def\Ecm{\ifmmode{E_{\mathrm{cm}}}\else{$E_{\mathrm{cm}}$}\fi}
\newc{\rts}{\sqrt{s}}
\newc{\rtshat}{\sqrt{\hat s}}
\newc{\gev}{\,GeV}
\newc{\mev}{~{\rm MeV}}
\newc{\tev}  {\mbox{$\;{\rm TeV}$}}
\newc{\gevc} {\mbox{$\;{\rm GeV}/c$}}
\newc{\gevcc}{\mbox{$\;{\rm GeV}/c^2$}}
\newc{\intlum}{\mbox{${ \int {\cal L} \; dt}$}}
\newc{\call}{{\cal L}}
\def \met  {\mbox{${E\!\!\!\!/_T}$}}

\newc{\ptmiss}{/ \hskip-7pt p_T}

\newc{\PT}{\mbox{$p_T$}}
\newc{\ET}{\mbox{$E_T$}}
\newc{\dedx}{\mbox{${\rm d}E/{\rm d}x$}}
\newc{\ifb}{\mbox{${\rm fb}^{-1}$}}
\newc{\ipb}{\mbox{${\rm pb}^{-1}$}}
\newc{\pb}{~{\rm pb}}
\newc{\fb}{~{\rm fb}}
\newc{\ycut}{y_{\mathrm{cut}}}
\newc{\chis}{\mbox{$\chi^{2}$}}

\def \jet(s){\emph{jet(s) }}

\newc{\mpl}{M_{\rm Pl}}
\newc{\mgut}{M_{GUT}}
\newc{\mw}{M_{W}}
\newc{\mweak}{M_{weak}}
\newc{\mz}{M_{Z}}

\newc{\OPALColl}   {OPAL Collaboration}
\newc{\ALEPHColl}  {ALEPH Collaboration}
\newc{\DELPHIColl} {DELPHI Collaboration}
\newc{\XLColl}     {L3 Collaboration}
\newc{\JADEColl}   {JADE Collaboration}
\newc{\CDFColl}    {CDF Collaboration}
\newc{\DXColl}     {D0 Collaboration}
\newc{\HXColl}     {H1 Collaboration}
\newc{\ZEUSColl}   {ZEUS Collaboration}
\newc{\LEPColl}    {LEP Collaboration}
\newc{\ATLASColl}  {ATLAS Collaboration}
\newc{\CMSColl}    {CMS Collaboration}
\newc{\UAColl}     {UA Collaboration}
\newc{\KAMLANDColl}{KamLAND Collaboration}
\newc{\IMBColl}    {IMB Collaboration}
\newc{\KAMIOColl}  {Kamiokande Collaboration}
\newc{\SKAMIOColl} {Super-Kamiokande Collaboration}
\newc{\SUDANTColl} {Soudan-2 Collaboration}
\newc{\MACROColl}  {MACRO Collaboration}
\newc{\GALLEXColl} {GALLEX Collaboration}
\newc{\GNOColl}    {GNO Collaboration}
\newc{\SAGEColl}   {SAGE Collaboration}
\newc{\SNOColl}    {SNO Collaboration}
\newc{\CHOOZColl}  {CHOOZ Collaboration}
\newc{\PDGColl}  {Particle Data Group Collaboration}

\def\issue(#1,#2,#3){{\bf #1}, #2 (#3)}
\def\iss(#1,#2,#3){{\bf #1} (#3) #2}
\def\issuenew(#1,#2,#3,#4){{\bf #1} (#2) {\bf no.#4}, #3}
\def\ASTR(#1,#2,#3){Astropart.\ Phys. \issue(#1,#2,#3)}
\def\AJ(#1,#2,#3){Astrophysical.\ Jour. \issue(#1,#2,#3)}
\def\AJS(#1,#2,#3){Astrophys.\ J.\ Suppl. \issue(#1,#2,#3)}
\def\APP(#1,#2,#3){Acta.\ Phys.\ Pol. \issue(#1,#2,#3)}
\def\JCAP(#1,#2,#3){Journal\ XX. \issue(#1,#2,#3)} 
\def\SC(#1,#2,#3){Science \issue(#1,#2,#3)}
\def\PRD(#1,#2,#3){Phys.\ Rev.\ D \issue(#1,#2,#3)}
\def\PR(#1,#2,#3){Phys.\ Rev.\ \issue(#1,#2,#3)} 
\def\PRC(#1,#2,#3){Phys.\ Rev.\ C \issue(#1,#2,#3)}
\def\NPB(#1,#2,#3){Nucl.\ Phys.\ B \issue(#1,#2,#3)}
\def\NPPS(#1,#2,#3){Nucl.\ Phys.\ Proc. \ Suppl \issue(#1,#2,#3)}
\def\NJP(#1,#2,#3){New.\ J.\ Phys. \issue(#1,#2,#3)}
\def\JP(#1,#2,#3){J.\ Phys.\issue(#1,#2,#3)}
\def\JPG(#1,#2,#3){J.\ Phys.\ G \issue(#1,#2,#3)}
\def\PL(#1,#2,#3){Phys.\ Lett. \issue(#1,#2,#3)}
\def\ZP(#1,#2,#3){Z.\ Phys. \issue(#1,#2,#3)}
\def\ZPC(#1,#2,#3){Z.\ Phys.\ C  \issue(#1,#2,#3)}
\def\PREP(#1,#2,#3){Phys.\ Rep. \issue(#1,#2,#3)}
\def\PRL(#1,#2,#3){Phys.\ Rev.\ Lett. \issue(#1,#2,#3)}
\def\MPL(#1,#2,#3){Mod.\ Phys.\ Lett. \issue(#1,#2,#3)}
\def\RMP(#1,#2,#3){Rev.\ Mod.\ Phys. \issue(#1,#2,#3)}
\def\SJNP(#1,#2,#3){Sov.\ J.\ Nucl.\ Phys. \issue(#1,#2,#3)}
\def\CPC(#1,#2,#3){Comp.\ Phys.\ Comm. \issue(#1,#2,#3)}
\def\IJMPA(#1,#2,#3){Int.\ J.\ Mod. \ Phys.\ A \issue(#1,#2,#3)}
\def\MPLA(#1,#2,#3){Mod.\ Phys.\ Lett.\ A \issue(#1,#2,#3)}
\def\PTP(#1,#2,#3){Prog.\ Theor.\ Phys. \issue(#1,#2,#3)}
\def\RMP(#1,#2,#3){Rev.\ Mod.\ Phys. \issue(#1,#2,#3)}
\def\NIMA(#1,#2,#3){Nucl.\ Instrum.\ Methods \ A \issue(#1,#2,#3)}
\def\EPJC(#1,#2,#3){Eur.\ Phys.\ J.\ C \issue(#1,#2,#3)}
\def\RPP (#1,#2,#3){Rept.\ Prog.\ Phys. \issue(#1,#2,#3)}
\def\PPNP(#1,#2,#3){ Prog.\ Part.\ Nucl.\ Phys. \issue(#1,#2,#3)}
\newc{\PRDR}[3]{{Phys. Rev. D} {\bf #1}, Rapid  Communications, #2 (#3)}

\def\PLB(#1,#2,#3){Phys.\ Lett.\ B  \issue(#1,#2,#3)}
\def\JHEP(#1,#2,#3){JHEP \issue(#1,#2,#3)}
\def\PRDnew(#1,#2,#3,#4){Phys.\ Rev.\ D \issuenew(#1,#2,#3,#4)}
\def\EPJCnew(#1,#2,#3){Eur.\ Phys.\ J.\ C \issuenew(#1,#2,#3)}
\def\JCAPnew(#1,#2,#3,#4){JCAP \issuenew(#1,#2,#3,#4)}

\def\gmin2{(g-2)_\mu}

\catcode`\@=11 

\def\vev#1{\left\langle #1\right\rangle}
\def\lsim{\mathrel{\mathpalette\@versim<}}
\def\gsim{\mathrel{\mathpalette\@versim>}}
\def\@versim#1#2{\vcenter{\offinterlineskip
    \ialign{$\m@th#1\hfil##\hfil$\crcr#2\crcr\sim\crcr } }}
\def\etal{{\em et. al.}}

\def\r2{\sqrt 2}
\def\beq{\begin{equation}}
\def\eeq{\end{equation}}
\def\beqn{\begin{eqnarray}}
\def\eeqn{\end{eqnarray}}
\def\sinW2{\sin^2\theta_W}

\def\mz2{M_{z}^2}
\def\c2b{\cos 2\beta}

\def\m#1{{\tilde m}_#1}

\def\mw#1{{\tilde m}_{\omega #1}}

\def\mz{M_Z}
\def\m0{m_0}
\def\mhalf{m_{\frac{1}{2}}}

\def\cb{\cos\beta}

\def\sec2w{sec^2\theta_W}

\def\gmin2{(g-2)_\mu}

\catcode`\@=11 

\def\vev#1{\left\langle #1\right\rangle}
\def\lsim{\mathrel{\mathpalette\@versim<}}
\def\gsim{\mathrel{\mathpalette\@versim>}}
\def\@versim#1#2{\vcenter{\offinterlineskip
    \ialign{$\m@th#1\hfil##\hfil$\crcr#2\crcr\sim\crcr } }}
\def\etal{{\em et. al.}}

\def\tb{\tilde b}

\def\tL{\tilde L}

\def \chonep{{\wt\chi_1}^{+}}
\def \chonem{{\wt\chi_1^-}}
\def \chonep2{{\wt\chi_2^+}}
\def \chonem2{{\wt\chi_2^-}}

\def \chonjpm{{\wt\chi_j}^{\pm}}
\def \chonepm{{\wt\chi_1}^{\pm}}

\def \mchonepm{m_{\chonepm}}

\def \chtwopm{{\wt\chi_2}^{\pm}}
\def \mchtwopm{m_{\chtwopm}}

\def \lstop{\wt{t}_{1}}

\def \lspi{\wt\chi_i^0}

\def \lspone{\wt\chi_1^0}
\def \mlspone{m_{\lspone}}
\def \lsptwo{\wt\chi_2^0}
\def \mlsptwo{m_{\lsptwo}}
\def \lspthree{\wt\chi_3^0}
\def \mlspthree{m_{\lspthree}}
\def \lspfour{\wt\chi_4^0}
\def \mlspfour{m_{\lspfour}}

\def\PL{Phys. Lett.}
\def\PRL{Phys. Rev. Lett.}

\def\PR{Phys. Rev.}

\def \lsptwo{\wt\chi_2^0}
\def \lspone{\wt\chi_1^0}
\def\lspi{wt\chi_i^0}
\def \chonem {{\wt\chi_1^\pm}}
\def \chargino1 {{\wt\chi_1^\pm}}
\def \chargino2 {{\wt\chi_2^\pm}}
\def \lstop{\wt{t}_{1}}
\def \ch2m {{\wt\chi_2^-}}
\def \lspi{\wt\chi_i^0}
\def \chonep {{\wt\chi_1^+}}

\begin{document}

\renewcommand*{\thefootnote}{\fnsymbol{footnote}}


\title{Multilepton signals of heavier electroweakinos at the LHC}


\vspace{5mm}

\author{Manimala Chakraborti}
\email{mani.chakraborti@gmail.com}
\affiliation{Bethe Center for Theoretical Physics \& Physikalisches Institut der Universit\"at Bonn, Nu{\ss}allee 12, 53115 Bonn, Germany}
\author{Amitava Datta}
\email{adatta\_ju@yahoo.co.in}
\affiliation{Fellow of Indian National Science Academy,\\
Bahadur Shah Zafar Marg, New Delhi - 110002, India}
\author{Nabanita Ganguly}
\email{nabanita.rimpi@gmail.com}
\affiliation{Department of Physics, University of Calcutta,\\
92 Acharya Prafulla Chandra Road, Kolkata - 700009, India}
\author{Sujoy Poddar}
\email{sujoy.phy@gmail.com}
\affiliation{Department of Physics, Diamond Harbour Women's University,\\
Sarisha, West Bengal - 743368, India}

\begin{abstract}\vspace*{10pt}

As sequel to a recent paper we examine the phenomenology of the full electroweakino sector of the pMSSM without invoking the adhoc but often employed assumption that the heavier ones are decoupled. We identify several generic models which illustrate the importance of the heavier electroweakinos and constrain them with the LHC $3l$ + ${E\!\!\!\!/_T}$ data. The constraints are usually stronger than that for decoupled heavier electroweakinos indicating that the LHC data is already sensitive to their presence. We also take into account the constraints from the observed dark matter relic density of the universe and precisely measured anomalous magnetic moment of the muon. Using the allowed parameter space thus obtained, we show that in addition to the conventional $3l$ + ${E\!\!\!\!/_T}$ signatures novel multilepton ($ml$) + ${E\!\!\!\!/_T}$ final states with $m > 3$, which are not viable in models with lighter electroweakinos only, can be observed before the next long shut down of the LHC.

\end{abstract}

\date{\today}




\maketitle

\renewcommand*{\thefootnote}{\arabic{footnote}}
\setcounter{footnote}{0}



\section{Introduction} 
The search for supersymmetry (SUSY) (For reviews on supersymmetry, see, {\it e.g.},\cite{susyrev1,susyrev2,susyrev3,susyrev4}-\cite{SUSYbooks1,SUSYbooks2}), the most well motivated extension of the Standard Model (SM) of particle physics, is in progress at the Large Hadron Collider (LHC) for the last few years \cite{atlastwiki},\cite{cmstwiki}. Unfortunately the experiments have yielded null results so far leading to limits on the masses of some supersymmetric partners - collectively known as the sparticles. 

In this paper we focus our attention on the electroweak (EW) sector of the supersymmetric standard model. In addition to novel LHC signatures this sector has several other important predictions. It is well known that the sparticles in this  sector may explain the origin of the observed dark matter (DM) relic density of the universe \cite{wmap,Ade}, for review see, e.g.,\cite{susydmrvw1,susydmrvw2,susydmrvw3,susydmrvw4,susydmrvw5},\cite{susydmrcnt1,susydmrcnt2,susydmrcnt3,susydmrcnt4,susydmrcnt5,susydmrcnt6,susydmrcnt7,susydmrcnt8,susydmrcnt9,susydmrcnt10} \footnote{However co-annihilation of strongly interacting sparticles with the LSP may produce the observed DM in the universe \cite{nlsp1,nlsp2}. More recent works can be found in \cite{coanni1,coanni2,coanni3,coanni4}}. In addition light sparticles belonging to this sector may also contribute to the anomalous magnetic moment of the muon ($a_{\mu}$) so that the alleged disagreement between this  precisely measured quantity \cite{muonexp1,muonexp2} and the SM  prediction (for a review see, e.g. , \cite{muonsm}) is significantly reduced due to contributions from virtual sparticles\cite{muonsusy}. In this context one must not forget the naturalness criterion (\cite{naturalness1,naturalness2,naturalness3}-\cite{naturalnessrvw}) one of the main motivations for invoking supersymmetry \footnote{For more recent ones see,e.g.,\cite{naturalsusy1,naturalsusy2}}. It is well known that the naturalness of a SUSY model crucially depends on the magnitude of the higgsino mass parameter $\mu$ \cite{susyrev1,susyrev2,susyrev3,susyrev4} which also belongs to the EW sector. The constraints on this parameter emerging from the LHC data and other observables can potentially test the naturalness of the models under consideration.  

The masses and other parameters belonging to the EW sector have been constrained by the $3l + \met$ searches by the LHC collaborations \cite{atlas3l,cms3l}. However, extracting these limits from the data is by no means straightforward. Ambiguities inevitably arise due to the fact that the SUSY breaking mechanism is yet to be discovered. As a result the masses of a plethora of sparticles and many other important parameters are essentially unknown. In order to simplify the analyses the number of unknown parameters contributing to a  particular SUSY signal is reduced by imposing additional assumptions which often turn out to be rather adhoc in nature. Obviously analyses reducing  such adhoc assumptions, as far as practicable, are desirable for drawing the conclusion on the viability of SUSY, a novel symmetry with many elegant features. Moreover such assumption may obfuscate novel signatures of SUSY which can show up at the LHC in near future as we shall show below. 

For example, the limits obtained from the searches for the electroweakinos (EWeakinos), the superpartners of the gauge and Higgs bosons, in the $3l + \met$ channel during LHC Run I \cite{atlas3l,cms3l} involved several restrictive assumptions regarding these sparticles. Our phenomenological analyses \cite{mc1,mc2} using ATLAS Run I data relaxed some of the above assumptions and showed that the constraints could be significantly weaker. However, all the above and several other recent phenomenological studies \cite{electroweakino1,electroweakino2,electroweakino3,electroweakino4,electroweakino5,electroweakino6} invoked an adhoc assumption that only a limited number of relatively light EWeakinos contribute to the $3l$ signal while the heavier ones are decoupled

The importance of the heavier EWeakinos and their LHC signatures were emphasized for the first time in \cite{ng}. It was illustrated  that the ATLAS  $3l$ data from Run I was already  quite sensitive to the masses of the heavier EWeakinos. Should this signal show up during LHC Run II models with both decoupled and non-decoupled EWeakinos must be included in attempts to decipher the underlying physics. More important, novel $ml + \met$  signatures with $m > 3$, which are not viable in many models with decoupled heavy EWeakinos, may show up before the next long shutdown of the LHC \cite{ng}.             

In this paper we  extend and complement \cite{ng} in several ways. First we make detailed study of LHC phenomenology in several scenarios briefly studied in \cite{ng} using a few benchmark points (BPs) only. Moreover, we delineate the allowed parameter space (APS) of several interesting models taking into account additional constraints like the observed dark matter (DM) relic density of the universe \cite{wmap,planck}, and the precisely measured anomalous magnetic moment of the muon ($a_{\mu}$) \cite{muonexp1,muonexp2} not considered in \cite{ng}. We also briefly comment on the naturalness (\cite{naturalness1,naturalness2,naturalness3}-\cite{naturalnessrvw}) of the models examined in this paper. Finally, we examine the prospect of observing the $ml + \met$ signature for $m \ge 3$ before the next long shutdown of the LHC. 

In Section \ref{models}, we briefly describe the models of EWeakinos studied in this paper and earlier works. In Section \ref{hvrewkno}, we illustrate the production and decay modes of the heavier EWeakinos with the help of benchmark points. In Section \ref{method}, we present the methodology followed for obtaining the main results of this paper. In Section \ref{newbound}, we analyse some of the models in Section \ref{models} using the constraints discussed in Section \ref{constraint} and identify the allowed parameter space in each case. In Section \ref{signal}, we illustrate the potentially observable $m l + \met$ signatures for $m \geq 3$ before the next long shutdown of the LHC. Finally, we conclude in Section \ref{conclude}.

\section{Models of non-decoupled heavier EWeakinos}    

In the  R-parity conserving minimal supersymmetric standard model (MSSM), the EW sector comprises of the following sparticles. The fermionic sparticles are the charginos ($\chonjpm$, $j= 1, 2$) and the neutralinos ($\lspi$, $i = 1 - 4$) - collectively called the EWeakinos. The masses and the compositions of these sparticles are determined by four parameters: the U(1) gaugino mass parameter $M_1$, the SU(2) gaugino mass parameter $M_2$, the higgsino mass parameter $\mu$  and tan $\beta$ - the ratio of the vacuum expectation values of the two neutral Higgs bosons. If no assumption regarding the  SUSY breaking mechanism is invoked the soft breaking masses $M_1$, $M_2$ and the superpotential parameter $\mu$ are all independent. Throughout this paper we take tan $\beta$ = 30 since relatively large values of this parameter give a better agreement with the $a_{\mu}$ data and ensure that the SM like Higgs boson has the maximum possible mass at the tree level. The indices j and i are arranged in ascending order of the masses. The stable, neutral lightest neutralino ($\lspone$), which is assumed to be the lightest supersymmetric particle (LSP), is a popular DM candidate. 

In the phenomenological MSSM (pMSSM) \cite{pmssm}, a  simplified  version of the MSSM, reasonable assumptions like negligible flavour changing neutral currents and CP violation are invoked to reduce the number of free parameters to 19. In this case the parameters belonging to the EWeakino sector introduced in the last paragraph are assumed to be real and the slepton mass matrices are assumed to be diagonal in the flavour basis.  All observables considered in this paper can be computed in this framework in a straightforward way.
    
The scalar sparticles are the $L$ and $R$ type sleptons and the sneutrinos. We assume L (R)-type sleptons of all flavours to be  mass degenerate with a common mass $\mslepl$ ($\mslepr$). Because of SU(2) symmetry the sneutrinos are mass  degenerate with L-sleptons modulo the D-term contribution. We neglect L-R mixing in the slepton sector. For simplicity we work in the decoupling regime (See {\it e.g.},\cite{djouadi}) of the Higgs sector of the MSSM with only one light, SM like Higgs boson, a scenario consistent with all Higgs data collected so far (See, {\it e.g.}, \cite{philip}).

The constraints on the EWeakino masses from the LHC searches are also sensitive to their compositions which are governed by the hierarchy among the parameters $M_1, M_2$ and $\mu$. Most of the existing analyses revolve around two broad scenarios listed below.\footnote{ We shall, however, briefly comment on other models as well.}

a) The Light Wino (LW) models : They are inspired by the simplified models employed by the LHC collaborations to interpret the $3l$ data. It is assumed that the lighter EWeakinos $\chonepm$ and $\lsptwo$ are purely wino and nearly mass degenerate while the $\lspone$ is bino dominated (\cite{atlas3l},\cite{cms3l},\cite{mc1}). 

This scenario can be easily realized in the pMSSM \cite{mc1}. Here  the wino dominated lighter EWeakinos, $\chonepm$ and $\lsptwo$, have nearly degenerate masses governed by the  parameter $M_2$ while the LSP ($\lspone$) could be bino dominated  with mass controlled by the U(1) gaugino mass parameter $M_1$ ($M_1 << M_2$). The heavier EWeakinos ($\chtwopm$, $\lspthree$ and $\lspfour$) are higgsino like with masses approximately equal to $\mu$, where $M_2 < \mu$. The somewhat adhoc assumption that these higgsino like sparticles are decoupled requires $M_2 << \mu$. This decoupling can be introduced in numerical computations by setting, e.g., $\mu = 2 M_2$ \cite{mc1}. 

However, computation of DM relic density immediately reveals that the results are not always consistent with the measured value if the compositions of the EWeakinos are exactly as stated above. For example, a glance at Fig 1b of \cite{mc1} indicates that these compositions are strictly realized in the parameter space with $\mlspone << \mchonepm$. On the other hand this parameter space is not consistent with the observed DM relic density of the universe. In fact the only parameter space allowed by the DM relic density constraint  (the upper red dotted line in this figure) corresponds to $\mlspone \approx \mchonepm$. In other words the DM constraint is satisfied if the $\lspone$ ($\chonepm$) though dominantly a bino (wino) has a sizable wino (bino) component in it. Thus consistency with  both LHC and DM constraints requires some admixture in the compositions of the EWeakinos. Similar conclusions follow for most of the LW models considered in \cite{mc1}. Moreover, the light wino models with typically large $\mu$ are also disfavoured by naturalness arguments. It is also worth noting that the wino like heavier EWeakinos in the LH model have relatively large cross-sections in spite of the suppression due to their large masses.

b) The Light Higgsino (LH) models : In this paper, following \cite{mc2,ng}, we mainly consider scenarios with higgsino dominated $\chonepm$, $\lsptwo$ and $\lspthree$ while the LSP is either bino dominated or a bino-higgsino admixture. The three lighter EWeakinos have closely spaced masses governed by $\mu$  while the $\lspone$ is either bino dominated  with mass controlled by  $M_1$ or a bino-higgsino admixture ($ M_1 \lsim \mu$). The two heavier EWeakinos ($\chtwopm$ and $\lspfour$) are wino like with masses approximately equal to $M_2$, where $M_2 > \mu$. However, the choice $M_2 = 2 \mu$ in \cite{mc2} effectively decouples these sparticles.

In summary all analyses \cite{atlas3l,cms3l,mc1,mc2} predating \cite{ng} invoked the adhoc assumption that the heavier EWeakinos are decoupled (i.e., $\mu >> M_2$ in case a) and $M_2 >> \mu$ in case b)).

In all models the trilepton rates  also depend sensitively on the hierarchy among the slepton and EWeakino masses. If the sleptons are lighter (heavier) than $\chonepm$ and $\lsptwo$, the leptonic Branching Ratios (BR) of these EWeakinos are typically large (small) yielding stronger (weaker) limits.

\label{models}

It may be recalled that the strongest lower limit  $\mchonepm > 800$ GeV for negligible LSP mass \cite{atlas3l} is obtained in the Light Wino with Light Left Slepton (LWLS) model. Here all L-sleptons masses are fixed at the arithmatic mean of $\mlspone$ and $\mchonepm$ which enhances the leptonic BRs of the decaying EWeakinos. All R-sleptons are assumed to be decoupled. In view of the above lower limit the heavier EWeakino masses are automatically restricted to be rather high so that they cannot significantly enhance the trilepton and other signals. If the L-sleptons are heavier than $\chonepm$, the bounds on $\mchonepm$ are much weaker ($\approx$ 350 GeV for negligible LSP masses). However, the production cross sections of the higgsino dominated heavier EWeakino pairs are in general suppressed as discussed in \cite{mc2}. Thus the cross sections of all multilepton signals from their cascade decays tend to be small. It is worth noting that the wino like heavier EWeakinos in the LH model have relatively large cross-sections in spite of the suppression due to their large masses.

The above discussions motivate us to  primarily focus on the LH type models with occasional comments on the LW models.

\subsection{The Compressed Light Higgsino Heavy Slepton (LHHS) model}

In the compressed LHHS model first considered in \cite{ng}, $M_1\simeq \mu$ with $M_2 >\mu$. This choice of parameters leads to a compressed lighter EWeakino spectrum where $\lspone$, $\lsptwo$, $\lspthree$ and $\chonepm$ are approximately mass degenerate and each has significant bino and higgsino components. The masses of the wino dominated heavier EWeakinos are determined by the free parameter $M_2$. Here the common slepton mass is set to be $m_{\tilde l_L} = m_{\tilde l_R} = (\mchonepm+\mchtwopm)/2 $ so that sleptons are always heavier than lighter EWeakinos. 

It is well known that if  the LSP is a bino-higgsino admixture the DM relic density constraint can be satisfied \cite{arkani}-\cite{naturalsusy1}. In addition this  model is worth  studying from the point of view of naturalness since $\mu$ is necessarily small. On the other hand since $\chonepm, \lsptwo$ and $\lspthree$ are nearly mass degenerate with $\lspone$, any signal stemming from the lighter EWeakino sector essentially consists of soft particles in the final state and are hard to detect. This tension is relaxed  if the  heavier EWeakinos ($\chtwopm$ and $\lspfour$) are relatively light. Observable multilepton signals from their cascade decays indeed  look promising \cite{ng}. This issue will be taken up in further details in the next section and Section \ref{newbound}. In the rest of this paper this model will be simply referred to as the compressed model.  
\label{complhhs}            

\subsection{The Light Higgsino Heavy Slepton (LHHS) model}

In this class of models \cite{mc2} the mass parameters $\mu < M_2$ whereas $M_1$ is taken to be the lightest. The mass of LSP is determined by $M_1$ and it is bino dominated. $\chonepm$, $\lsptwo$ and $\lspthree$ are higgsino dominated  with closely spaced masses controlled
by $\mu$ . The wino like heavier EWeakinos - $\chtwopm$ and $\lspfour$- have masses controlled by $M_2$. For numerical results  we set $M_2 =1.5 \mu$. Both left and right sleptons masses are taken to be midway between $\mchonepm$ and $\mchtwopm$. Consequently, only heavier EWeakinos can decay directly into sleptons with significant BRs. Although $\chtwopm, \lspfour$ have suppressed production cross-sections compared to the lighter EWeakinos, they have  moderately large lepton yield thanks to their cascade decays involving sleptons, lighter EWeakinos, $W$ and $Z$ bosons all of which can decay leptonically. One can, therefore, expect to get sizable multilepton ($4l$ and $5l$) signals from their cascade decays in additional to the conventional $3l$ final states. 
\label{lhhs}

\subsection{The Light Higgsino light Left Slepton (LHLS) model}

This model is the same as the previous one except that the masses of the L-sleptons are chosen to lie midway between $\mlspone$ and $\mlsptwo$/$\mchonepm$ whereas R-slepton masses are set at 2 TeV. Due to leptons coming from both lighter and heavier EWeakinos, the bounds on $\mchonepm$ or $\mlsptwo$ get stronger. Moreover, multilepton signals are also copiously produced.  
\label{lhls}

\subsection{The Light Mixed  light Left Slepton (LMLS) model}

In all mixed models EWeakinos except the LSP are admixtures of higgsino and wino components and have closely spaced masses (i.e., $\mu \simeq M_2$). The LSP is bino dominated with mass controlled by $M_1$. In the LMLS model the left sleptons masses are kept midway between $\lspone$ and $\chonepm$ whereas right sleptons are decoupled with masses $\simeq 2$ TeV.
\label{lmls}

\section{Production and decay modes of EWeakinos in different models} 

Since the production and decay modes of the heavier EWeakinos have not been discussed in recent literature we discuss main features in this section. It has already been pointed out that the production cross-section of lighter EWeakino pairs are quite sensitive to their compositions  (see \cite{mc2} Table 2). For fixed $\mlspone$ and $\mchonepm$, it is largest in the wino model, smallest in the higgsino model and intermediate in case of the mixed scenario. This result can be readily generalised to the heavy EWeakinos. Naturally the production cross section of  $\chtwopm-\lspfour$ pair in the Higgsino model is suppressed by their large masses. However, this suppression is to some extent compensated since they are wino like. In contrast the production cross sections of the lighter EWeakinos suffer suppression due to their higgsino like composition. At the production level the lighter and heavier EWeakino yields could, therefore, be quite compitative.  

In Table \ref{tab1} we present the cross-sections of different EWeakino pairs for representative points in models described in the previous section. The benchmark points P1 - P4 correspond to the compressed, LHHS, LHLS and LMLS models respectively. The total production cross section of only lighter EWeakino pairs (all EWeakino pairs involving at least one heavy EWeakino) is denoted by $\sigma(p p \ra$ LEWs) ($\sigma(p p \ra$ HEWs)). Comparing the two cross-sections it follows that in all cases the heavier EWeakinos in non-decoupled scenarios significantly contribute to the total EWeakino production in spite of their large masses. 

\begin{table}[H]
\begin{center}
\begin{tabular}{|c||c|c|c|c|}
\hline
\hline
Masses & \multicolumn{4}{c|}{Models} \\
\cline{2-5}
and &P1  &P2 &P3 &P4 \\
Cross-sections &(Compressed) &(LHHS) &(LHLS) &(LMLS) \\
\cline{1-5}
$M_1$ &186 &105 &249 &321 \\
$\mu$ &191 &270 &300 &401 \\
$M_2$ &350 &405 &450 &382 \\
$\mlspone$ &151 &100 &231 &304 \\
$\mlsptwo$ &198  &262 &304 &359 \\
$\mlspthree$ &213 &281 &311 &412 \\
$\mlspfour$ &389 &447 &491 &467 \\
$\mchonepm$ &178 &260 &291 &350 \\
$\mchtwopm$ &389 &447 &491 &465 \\
\hline
$\sigma(p p \ra$ LEWs) &621.9 &299.5 &165.8 &72.94  \\
$\sigma(p p \ra$ HEWs) &147.1 &81.4 &52.0 &83.95 \\
$\sigma_{tot}$  &768.9 &380.9 &217.8 &156.9 \\
\hline
\hline

\end{tabular}
\end{center}
\caption{Mass parameters, physical masses and production cross-sections of EWeakinos for four representative points in different models introduced in Section \ref{models}. See the text for the details. All masses and mass parameters are in GeV. Cross-sections are in $fb$.}
\label{tab1}
\end{table}

In Table \ref{tab2} we present the BRs of all EWeakinos in scenarios P1 - P4. In Table \ref{tab3} we have shown the effective cross-section defined as ($\sigma \times$ BR)$_{ml}$ of  multilepton channels with m = 3,4,5. Here LEW (HEW) refers to the contribution of lighter EWeakino pairs only (pairs with at least one heavier EWeakino) to a particular signal.

\begin{table}[H]
\begin{center}
\begin{tabular}{|c|c|c|c|c|c||c|c|c|c|}
\hline
\hline
Decay & \multicolumn{4}{c||}{Branching Ratio} &Decay &\multicolumn{4}{c|}{Branching Ratio} \\
\cline{2-5} \cline{7-10}
Modes &P1  &P2 &P3 &P4  &Modes &P1 &P2 &P3 &P4 \\
\cline{1-10}
\hline

$\chtwopm \ra \snu l^{\pm}$ &0.17 &0.14 &0.24 &0.24 &$\lspfour \ra \lspi Z$ &0.12 &0.11 &0.06 &- \\
$\ra \snu_{\tau_1} \tau^{\pm}$ &0.08 &0.07 &0.12 &0.15 &$\ra \chonepm W^{\mp}$ &0.30 &0.34 &0.17 &0.24 \\
$\ra \tilde{l}^{\pm}_L \nu$ &0.16 &0.14 &0.24 &0.28 &$\ra \lspi h$ &0.09 &0.10 &0.05 &0.01 \\
$\ra \tilde{\tau}^{\pm}_{1,2} \nu_{\tau}$ &0.08 &0.07 &0.12 &0.14 &$\ra \tilde{l}^{\pm}_L l^{\mp}$ &0.14 &0.11 &0.22 &0.16 \\
$\ra \chonepm Z$ &0.14 &0.16 &0.08 &0.08 &$\ra \tilde{\tau}^{\pm}_{1,2} \tau^{\mp}$ &0.07  &0.06 &0.11  &0.12 \\
$\ra \lspi W^{\pm}$ &0.27 &0.31 &0.15 &0.11 &$\ra \snu \nu$ &0.28 &0.24 &0.39 &0.47 \\
$\ra \chonepm h$ &0.09 &0.10 &0.17 &- & & & &  & \\
\cline{1-10}
$\chonepm \ra \lspone q \bar{q'}$ &0.66  &- &- &- &$\lspthree \ra \lspone q \bar{q}$ &0.02 &- &- &- \\
$\ra \lspone l^{\pm} \nu$ &0.22 &- &- &&$\ra \lspone l^+ l^-$ &0.02 &- &- &  \\
$\ra \lspone \tau^{\pm} \nu_{\tau}$ &0.11 &- &- &&$\ra \lspone \tau^+ \tau^-$ &0.03 &- &- &-\\
$\ra \lspone W^{\pm}$ &- &1.0 &- &&$\ra \lspone \nu \bar{\nu}$ &0.02 &- &- &- \\
$\ra \snu l^{\pm}$ &- &- &0.36 &0.45 &$\ra \chonepm q \bar{q}$ &0.60 &- &- &-   \\
$\ra \snu_{\tau_1} \tau^{\pm}$ &- &- &0.52 &0.28 &$\ra \chonepm \nu l^{\mp}$ &0.20 &- &- &- \\
$\ra \tilde{l}^{\pm}_L \nu$ &- &- &0.08 &0.18&$\ra \chonepm \nu_{\tau} \tau^{\mp}$ &0.10 &- &- &-  \\
$\ra \tilde{\tau}^{\pm}_1 \nu_{\tau}$ &- &- &0.04 &0.09 &$\ra \lspi Z$ &- &0.91 &- &0.44  \\
\cline{1-5}
$\lsptwo \ra \lspone q \bar{q}$ &0.63 &- &- &- &$\ra \lspi h$ &- &0.09 &- &- \\
$\ra \lspone l^+ l^-$ &0.06 &- &- &- &$\ra \tilde{l}^{\pm}_L l^{\mp}$ &- &- &0.01 &- \\
$\ra \lspone \tau^+ \tau^-$ &0.04 &- &-  &- &$\ra \tilde{\tau}^{\pm}_1 \tau^{\mp}$ &- &- &0.84 &0.49  \\
$\ra \lspone \nu \bar{\nu}$ &0.20 &- &- &- &$\ra \snu \nu$ &- &- &0.15 &0.06  \\
$\ra \lspone Z$ &- &0.28 &-  &- &$\ra \chonepm W^{\mp}$ &- &- &- &-  \\
$\ra \lspone h$ &- &0.72 &- &- &- & & & &   \\
$\ra \tilde{l}^{\pm}_L l^{\mp}$ &- &- &0.55 &0.45 & & & & & \\
$\ra \tilde{\tau}^{\pm}_1 \tau^{\mp}$ &- &- &0.45 &0.26 & & & && \\
$\ra \snu \nu$ &- &- &- &0.29 & & & & &  \\
\hline
\hline
\end{tabular}
\end{center}
\caption{Different decay modes of EWeakinos along with their BRs for four representative points as given in Table \ref{tab1}.}
\label{tab2}
\end{table}

\newpage

In the  compressed model (P1), $\sigma$ is much larger for lighter EWeakinos than that for the heavier EWeakinos as expected. But since the  sleptons are heavier than the lighter EWeakinos the latter cannot decay directly into sleptons which eventually yields leptonic states with large BRs. However, leptonic decays mediated by virtual $W$ or $Z$ bosons are allowed. Small leptonic BR of the gauge bosons suppresses the leptonic signals if only the lighter EWeakinos are considered. The situation, however, completely changes if one takes into consideration the contributions of heavier EWeakino decays. They can decay either directly into sleptons or leptons can come from lighter EWeakino, $W$ or $Z$ mediated  processes. This enhances the multilepton signals quite a bit. This is illustrated in Table \ref{tab3} where we have shown the effective cross-section for different  multilepton channels. In the compressed model the leptonic decays of the lighter EWeakinos inevitably lead to soft leptons because of the small energy release. The apparently non-vanishing cross-sections in Table \ref{tab3} are drastically suppressed when appropriate cuts requiring hard leptons in the signal are imposed. This will be discussed in details in a later section.

\begin{table}[H]
\begin{center}
\begin{tabular}{|c|c|c|c|c|}
\hline
\hline
($\sigma \times$BR)$_{3l}$  &P1  &P2 &P3 &P4  \\
\hline
LEWs &9.36 &2.41 &18.25 &20.5 \\
\hline
HEWs &64.2 &4.85 &6.23 &6.46 \\
\hline
\hline
($\sigma \times$BR)$_{4l}$  &P1  &P2 &P3 &P4  \\
\hline
LEWs &0.212 &0.113 &0.116 &- \\
\hline
HEWs &20.2 &0.764 &0.661 &0.725 \\ 
\hline
\hline
($\sigma \times$BR)$_{5l}$  &P1  &P2 &P3 &P4  \\
\hline
LEWs &- &0.008 &- &- \\
\hline
HEWs &4.81 &0.134 &0.137 &0.118 \\
\hline
\hline

\end{tabular}
\end{center}
\caption{Effective cross-sections, namely $\sigma \times$BR, for three different signals - $3l$, $4l$ and $5l$ for four representative points in Table \ref{tab1}. See text for the detail. All the cross-sections are in $fb$.}
\label{tab3}
\end{table}

Next we come to the LHHS model (P2). Again the production cross-section of the lighter EWeakinos dominate over that of the heavier ones. However, heavier EWeakinos have larger leptonic BRs for reasons discussed in the last paragraph. From Table \ref{tab3}, one can see the relative contributions of lighter and heavier EWeakinos to multilepton signals. For the $3l + \met$ channel both of them have non-negligible contributions although bulk of the events come from the heavier ones. However, the heavier EWeakino contributions to $4l$ and especially $5l$ signals are much larger.

For the LHLS (P3) and LMLS (P4), the $3l$ signal is dominated by the lighter EWeakino contributions. However, the $4l$ and $5l$ signals are essentially due to the heavier ones.

We conclude this section with the important message that the multilepton signals in the compressed scenario completely depend on heavier EWeakinos. For other models although both heavy and light EWeakinos can contribute significantly to $3l + \met$, final states with more than three leptons essentially come from  the heavier ones. 

\label{hvrewkno}

\section{The methodology} 
\noindent
In this section we briefly describe the procedure for constraining the models presented in Section \ref{models}. We also outline the generator level simulation of different LHC signals using PYTHIA (v6.428) \cite{pythia} and the methods for scanning the parameter space. 

\label{method}

\subsection{The Constraints}
\noindent
We have used three major constraints involving relatively small theoretical/experimental uncertainties as listed below.

\begin{itemize} 

{\item We use the ATLAS  trilepton data \cite{atlas3l} on $\chonepm-\lsptwo$ searches from LHC Run I. The correlated constraints on LSP and slepton masses as given by the ATLAS slepton search data \cite{sleprun1} are also taken into account in models with light sleptons. We also require the lighter Higgs boson mass $m_h$ to be in the interval $122<m_h<128$ GeV around a central value of 125 GeV \cite{higgs1,higgs2}. This is achieved by choosing judiciously  the third generation trilinear soft breaking term - $A_t$, the CP odd Higgs mass - $M_A$ and the lighter top squark mass which is chosen to be 1.5 TeV. The window of 3 GeV mainly reflects the theoretical uncertainty \cite{higgsuncertainty1,higgsuncertainty2,higgsuncertainty3,higgsuncertainty4,higgsuncertainty5,higgsuncertainty6} in computing the Higgs mass in a typical SUSY scenario. The heavier Higgs bosons are assumed to be decoupled. It may be recalled that the BRs of the unstable EWeakinos depend on $m_h$. }

{\item The precise measurement of the muon anomalous magnetic moment ($a_{\mu} ={1\over2} {(g-2)_{\mu}}$) \cite{muonexp1,muonexp2} plays an important role in studying new physics. The experimental value of $a_{\mu}$ denoted by $a^{exp}_{\mu}$ differs significantly from the SM prediction $a^{SM}_{\mu}$ (see, e.g., \cite{muonsm}). This large deviation strongly hints for new physics beyond SM. There are three parts in $a^{SM}_{\mu}$ -  a part from pure quantum electrodynamics, a part from electroweak contributions and the hadronic contributions. The SUSY contribution to $a_{\mu}$, namely $a^{SUSY}_{\mu}$ becomes large when the charginos, neutralinos and smuons are relatively light \cite{muonsusy} \footnote{For more recent ones see, e.g., \cite{gminus21,gminus22,gminus23,gminus24,gminus25,gminus26}} and it scales with tan$\beta$. Thus one can constrain the SUSY parameter space with the measured upper and lower bounds on $\Delta a_{\mu}= a^{exp}_{\mu}-a^{SM}_{\mu}$ given by \cite{muonsusy} :
\be
\Delta a_{\mu} = a^{exp}_{\mu}-a^{SM}_{\mu} = (29.3\pm9.0)\times 10^{-10}
\ee
The computation of  $a^{SUSY}_{\mu}$ may be found in refs. \cite{susyg-2A1,susyg-2A2,susyg-2B1,susyg-2B2,susyg-2B3,endo}. It should be noted that the Higgs mass at 125 GeV and stringent lower bounds on the masses of the strong sparticles from LHC data strongly disfavour the models like mSUGRA with strong assumptions on soft breaking parameters  in the light of $a_{\mu}$ data \cite{baer-prannath1,baer-prannath2}. However, non-universal gaugino mass models can still resolve the $a_{\mu}$ anomaly within the said range of $\Delta a_{\mu}$ \cite{Nonunivg-21,Nonunivg-22,Nonunivg-23,Nonunivg-24}. This sensitivity shows that the $a_{\mu}$ data can indeed constrain the  slepton and EWeakino masses which are free parameters in the pMSSM. 

Since the SM is consistent with the measured $a_{\mu}$ at the 3$\sigma$ level we require that the models under scrutiny yield a better agreement, say, at the level of $\lsim 2 \sigma$.}

\item In this analysis we also impose the  constraint from the measured DM relic density of the universe \cite{wmap,planck}. The $3\sigma$ limit which we use in this work is :\\
\be
0.092<\Omega_{\tilde \chi}h^2 <0.138
\ee

Apparently this limit is significantly relaxed than the latest experimental limit  $0.1199 \pm 0.0022$ \cite{Ade}. This is due to the fact that the above range includes, in addition to the  experimental errors, estimated theoretical errors discussed in the literature \cite{Kozaczuk:2013spa}-\cite{Bertone}. A range similar to the above has been used, e.g., in \cite{badziak}. Recently it has also been noted that QCD corrections to neutralino annihilation and coannihilation channels involving quarks may introduce sizable uncertainties due to the choice of the QCD scale \cite{Harz}. In view of the above discussions a more conservative handling of the constraint seems to be justified.

\end{itemize}

There are other measurements related to DM which are often used to constrain the EWeakino sector of the pMSSM. However, theoretical and experimental uncertainties make them less stringent compared to the ones presented above. In the following we briefly summarize them (see the references given below) and indicate how we  have used them in our analysis.

The direct detection experiments measure the DM - nucleon scattering \cite{luxrcnt} cross sections. Since no scattering has been observed many models of DM have been constrained. It should be borne in mind that the theoretical prediction for the spin independent scattering cross-section ($\sigma_{SI}$) crucially depends on the value of the DM relic density ($\rho_E$) at the detector (i.e., in the neighbourhood of the earth). There are standard astrophysical methods for measuring the local density of DM ($\rho_L$) which is an average over a volume having a radius of typically a few hundred  parsecs (pcs) with the sun at the center \cite{read}. This volume though cosmologically small is  huge in the terrestrial length scale. The measured  central values of $\rho_L$ lies in the range 0.023 - 0.85 GeV $pc^{-3}$(see table 4 of \cite{read}). However, due to large errors the measured values are comparable with much smaller $\rho_{L}$. However, there is an even bigger source of uncertainty. Getting $\rho_E$ from $\rho_L$ involves an extrapolation over many orders of magnitudes which is mainly done by simulation. According to \cite{read} the Dark Matter Only (DMO) simulations indicate that $\rho_{E}$ and $\rho_{L}$ may not be very different. The situation, however, is further complicated by the presence of significant amount of ordinary baryonic matter in the solar system and its
possible impact on $\rho_E$. According to \cite{read} the  present techniques cannot predict a reliable $\rho_E$. On the other hand the global measure of dark matter density $\rho_G$ obtained  from the rotation curve of our galaxy typically have smaller errors (see Table 4 \cite{read}). But they are based on the strong assumption that the galactic halo is spherically symmetric which may not be realistic \cite{read}. 

It is well known that there is another way of evading the direct detection limits. For a mixed DM, which is the case for most of the models studied in this paper, it is possible that $\lspone - \lspone - h$ coupling is significantly suppressed in certain regions of the parameter space (the so called `blind spots') due to cancellation between different contributions to this coupling \cite{cheung}. As a result the theoretical prediction for $\sigma^{SI}$ may be further suppressed making it compatible with the direct detection data. In this paper, however, we do not examine the implications of the blind spots numerically.   

 There are other theoretical/ experimental  uncertainties in $\sigma^{SI}$ (e.g., the uncertainties in the form factors of LSP - quark scattering) as discussed in section 2.3 of \cite{mc1} where reference to the original works may be found. Taking all these into account the total uncertainty in the upper limit of the DM - nucleon cross-section could be  one order of magnitude or even larger. The spin independent DM - nucleon scattering cross-section $\sigma^{SI}$ has been computed in several LH models \cite{mc2}. It was argued that the models studied were compatible with  the then LUX data \cite{lux} on the upper bound on $\sigma^{SI}$  as a function of the DM mass provided allowance was made for the large uncertainties discussed above. However more recent LUX data (\cite{luxrcnt}) have imposed much stronger constraints on $\sigma^{SI}$. This  data  taken at its face value imposes strong lower limits on  the DM mass ($\mlspone$ ) which would make some models considered in this paper uninteresting in the context of the ongoing LHC searches. We shall come back to this issue in the next section when we consider different models.

There are interesting attempts to link several anomalies in astrophysical observations with annihilation of DM (see e.g., \cite{dmrvw1}-\cite{dmrvw2}). Many of the reported  signals are not free from ambiguities because of the uncertainties in the estimation of the astrophysics backgrounds. Moreover some of the reported results have not been confirmed by the subsequent experiments.  However, even if a few of the reported signals survive the test of time the underlying theory/theories  must have multiple DM particles with masses in the  range of a few keV to a few TeV. The pMSSM with a single DM candidate can  then at best be a part of a bigger scenario having multiple DM candidates. We, therefore, do not include these constraints in our analysis.

\label{constraint}

\subsection{The Simulation}
\noindent

We follow the analysis by ATLAS Collaboration for Run I data where they introduced 20 signal regions (SR) each characterized by a set of cuts \cite{atlas3l}. In Tables 7 and 8  of \cite{atlas3l} the model independent upper limit on $N_{BSM}$ at 95\% CL for each SR is shown. A point in the parameter space of any  model is said to be excluded if the corresponding $N_{BSM}$ exceeds the upper bound for atleast one of the 20 SRs in \cite{atlas3l}. We validate our simulation (\cite{mc1} and \cite{arghya}) and draw the exclusion contours for the models under consideration using PYTHIA. However, in this work we draw the exclusion contours considering the productions of all combinations of EWeakinos - heavy as well as light. For computing the NLO cross-sections of EWeakino production we use PROSPINO 2.1 \cite{prospino}.

We also simulate multilepton signals ($\ge 3l$) at 13 TeV LHC using PYTHIA which is described in later sections. The judicious choice of cuts are made in order to tame down the potentially dangerous background in each case. The relevant background events are generated using ALPGEN (v 2.1)\cite{alpgen} with MLM matching \cite{mlm1,mlm2} and then the generated events are fed to PYTHIA for showering and hadronization. Reconstruction of jets is made following the anti-$k_t$ algorithm \cite{antikt} by interfacing PYTHIA with FASTJET \cite{fastjet} with $R = 0.4$. The reconstructed jets are required to have $P_{T} > 20$ GeV and $|\eta|<2.5$. Also all the leptons (e and $\mu$) in the final state must have $P_{T} > 10$ GeV and $|\eta|<2.5$. In addition to that, each of them is required to pass isolation cuts as defined by the ATLAS/CMS collaboration \cite{atlas3l}-\cite{cms3l}. We use these selection cuts for all our analyses in this work. We use CTEQ6L \cite{cteq6l} parton density function (PDF) in our simulations of all signals.

\label{simulation}

\subsection{Scanning the parameter space} 

We use the following pMSSM parameters throughout this study. The squark masses belonging to the first two generations, $M_A$ and $M_3$ are set at a large value of 2 TeV. Note that these parameters do not have any effect on EW sector which is our main concern in this paper. The trilinear coupling $A_t$ is set at - 2 TeV so that the Higgs mass is consistent with the measured value. All other trilinear couplings namely $A_b$, $A_{\tau}$, $A_u$, $A_d$, $A_e$ are assumed to be zero. $M_1$, $M_2$ are scanned in the interval (see Section \ref{models}) while $\tan \beta$ is fixed at a high value 30.  The parameter $\mu$ is chosen to ensure the characteristics of each model. Thus in the compressed model we take $\mu = 1.05 M_1$ \footnote{The consequences for other choices will be discussed in the next section.}. The choices for the other models will be given in the next section. The slepton masses are fixed as discussed in  Section \ref{models}. The SM parameters considered are $m^{pole}_t = 175$ GeV, $m^{\bar{MS}}_b = 4.25$ GeV, $m_{\tau} = 1.77$ GeV and $M_Z = 91.18$ GeV. We consider only positive sign of $\mu$ in our work.

In this work we use SuSpect(v 2.41) \cite{suspect} for obtaining mass spectra and for evaluating $a^{SUSY}_{\mu}$. We compute the decay BRs of sparticles using SUSYHIT \cite{susyhit}. Calculations of relic density and $\sigma^{SI}$ are done with micrOMEGAs (v3.2) \cite{micromega3}.

\section{Constraining models with non-decoupled heavier EWeakinos}

In this section we delineate the APS of the models described in Section \ref{models} using the constraints discussed in Subsection \ref{constraint}. 

\label{newbound}

\subsection{Compressed LHHS Model}

As already discussed the compressed LHHS or simply the compressed Model model with  $\mu \approx M_1$  and approximately mass degenerate LSP and other lighter EWeakinos is attractive from the point of view of DM relic density and naturalness (\cite{wmap},\cite{planck},\cite{naturalness1,naturalness2,naturalness3},\cite{naturalnessrvw}). On the other hand the heavier EWeakinos - if  non-decoupled - can  lead to viable multilepton signals at the LHC as we shall show in this and the next section. 

\begin{figure}[h]
\centering
\includegraphics[width=0.5\textwidth]{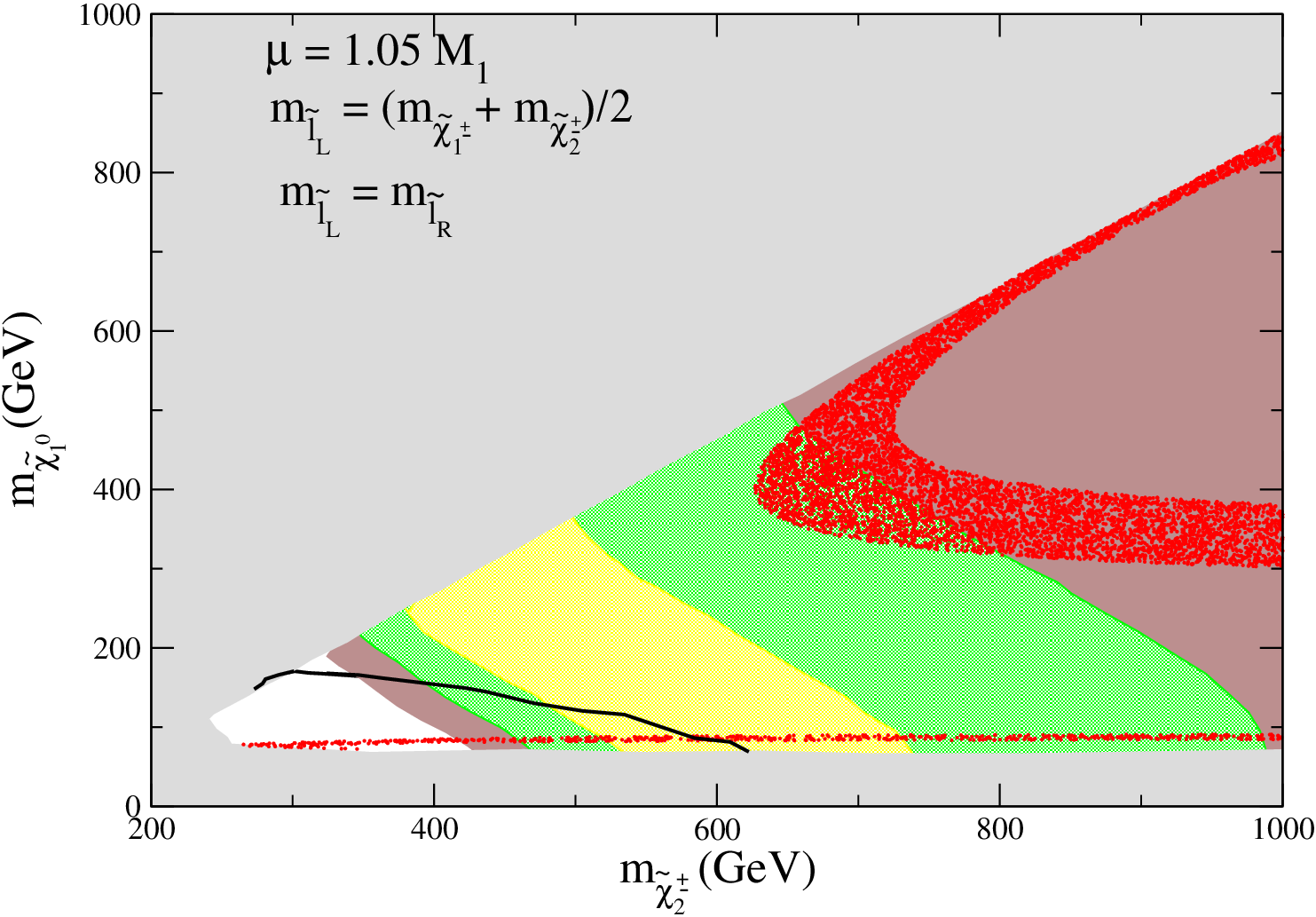}

\caption{The black line represents the exclusion contour in $\mchtwopm-\mlspone$ plane in the compressed model obtained by our simulation using ATLAS data \cite{atlas3l}. The brown, green and yellow regions in the parameter space are consistent with $a_{\mu}$ data at $3\sigma$, $2\sigma$ and $1\sigma$ level respectively. The red points satisfy WMAP/PLANCK data of DM relic density.}
\label{fig1}
\end{figure}

In Fig. \ref{fig1} we display the constraints in the $\mlspone - \mchtwopm$ plane  of the compressed model (see Section \ref{complhhs}). 
The black  exclusion contour, which is the first published constraints on $\mchtwopm$ \cite{ng}, represents the  bounds from the ATLAS $3l + \met$ data. For $\mlspone \simeq 80$ GeV, there is a strong bound $\mchtwopm > 610$ GeV. The LSP mass can not be lowered further due to the LEP bound $\mchonepm$ $> 103$ GeV \cite{lepsusy}. However, note that for $\mlspone \ge 170$ GeV, there is no bound on $\mchtwopm$. On the whole it follows that the sensitivity to the $3l$ data increases significantly if the heavier EWeakinos are non-decoupled. The exclusion gets weaker for $\mchtwopm <300$ GeV, since the higgsino component in $\chtwopm$ dominantes over the wino component. As a result the production cross-section is suppressed. \\

The dominant contribution to $a_{\mu}$ comes from chargino - sneutrino loop. Although the loop involving $\chonepm$ dominates, the loop contribution of $\chtwopm$ (which is almost half of the former) comes with an opposite sign. Their combined effect helps to get the $a_{\mu}$ value in the right ballpark. Different coloured regions in Fig. \ref{fig1} (see the figure caption) represent different levels of agreement between the model predictions and the data. This also indicates that over a fairly large parameter space  better agreement than the SM is obtained.

There are several distinct regions consistent with WMAP/PLANCK constraints in Fig. \ref{fig1}. The main DM producing mechanism for the upper branch of the parabola like region is LSP - $\chonepm$ coannihilation. Some  contributions also come from LSP pair annihilations into $W^+  W^−$ mediated by $\chonepm$ and $\chtwopm$  although they are  subdominant due to relatively large $\mchonepm$. The lower branch of the parabola arises mainly due to $\chonepm$ mediated LSP pair annihilations into $W^+ W^-$, into $ZZ$ through $\lsptwo$ or $\lspthree$ and $t \bar{t}$ through virtual $Z$ exchange. Here LSP - $\chonepm$ coannnihilation is small. In the lowest band at fixed $\mlspone$, LSP pair annihilations into $W^+ W^−$ via  $\chonepm$ is the main process. However, annihilation into $f \bar f$ final states has a non negligible contribution. Note that, there is no region where DM production proceeds via $h$ or $Z$ resonance since the LSP masses required for these processes are not allowed by the  mass bound on $\chonepm$ from LEP. 

It may be noted that in the parabola like region $\mchtwopm$ is rather high ($ > 600$ GeV) while in the lower band smaller  values of $\mchtwopm$ are ruled out by the LHC constraints. Such high values of $\mchtwopm$ tend to suppress the multilepton signals below the observable level for integrated luminosities expected to accumulate before the next long shutdown (see the next section). \\

\begin{figure}[h]
\centering
\includegraphics[width=0.5\textwidth]{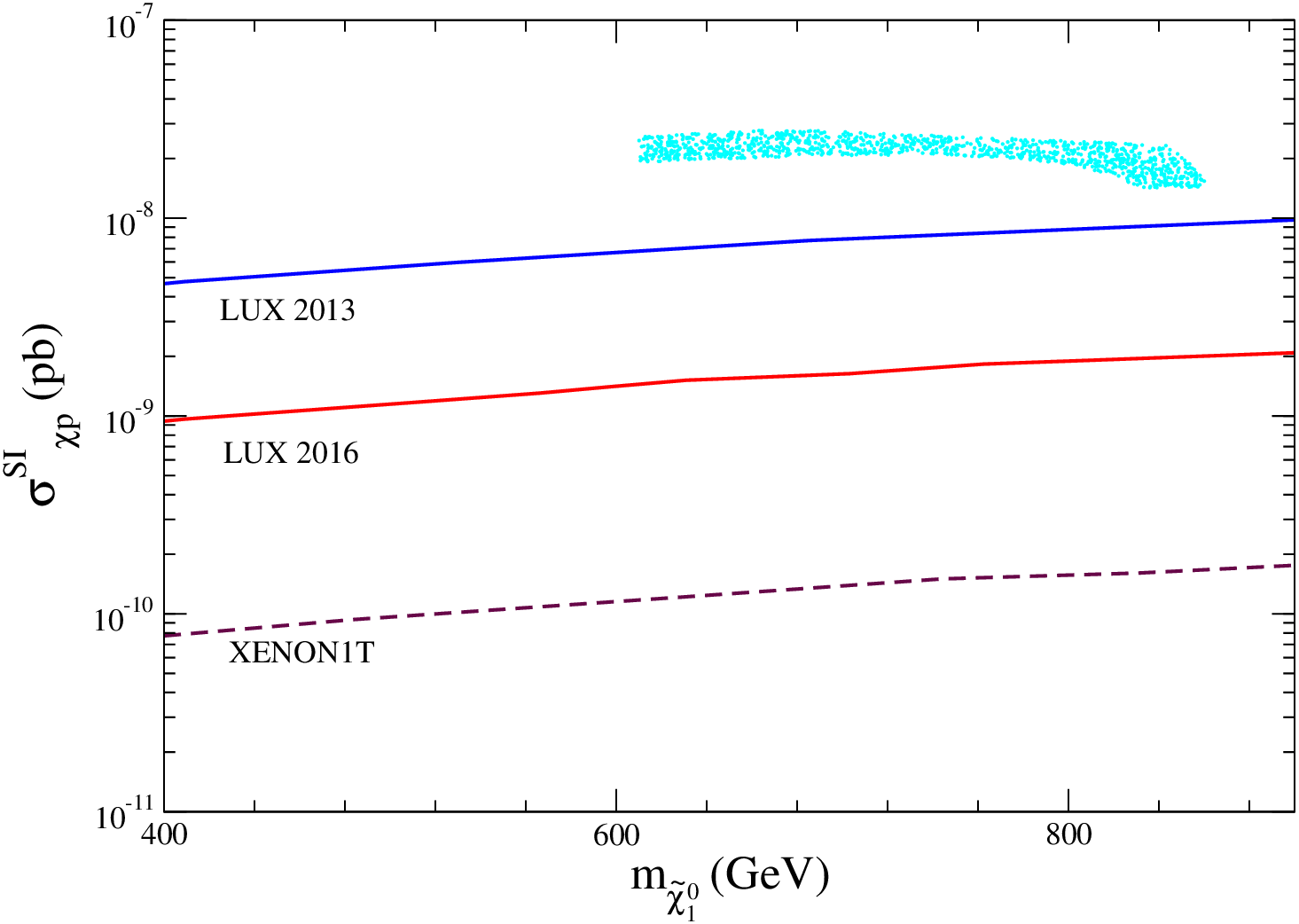}

\caption{Direct detection results for compressed model. LUX and XENON1T results are shown as solid red and dashed magenta lines respectively. Blue points satisfy WMAP/PLANCK data, LHC constraints and $a_{\mu}$ up to the level of $2\sigma$.}
\label{fig2}
\end{figure} 

 The DM relic density constraint is compatible  with lighter  $\chtwopm$ if extreme compression, represented  by the choice $\mu = 1.05 M_1$
is relaxed to some extent. For extreme compression the mass gap between LSP and $\chonepm$ is approximately 30 GeV irrespective of the specific choice of $M_1$. As a result both LSP - LSP annihilation and LSP - $\chonepm$ co-annihilation are pronounced yielding the observed DM relic density. For $\mchtwopm < 600$ GeV the wino component in LSP is  larger which further enhances the LSP annihilation rate leading to  under abundance of DM relic density. If the compression is relaxed to some extent by increasing the ratio $\mu/~M_1$ the WMAP/PLANCK constraint can be satisfied for lower $\mchtwopm$ favourable for multilepton signals. We will illustrate the impact of the modified scenario on multilepton signals in the next section with the help of BPs.

It has already been noted that the recent LUX upper bounds on $\sigma^{SI}$ \cite{luxrcnt} are in conflict with bino-higgsino DM \cite{bnohgsndm} with low masses. We plot in Fig. \ref{fig2} the prediction for $\sigma^{SI}$ in the compressed model for the APS in Fig. \ref{fig1}. It follows that $\sigma^{SI}$ is typically $\simeq 10^{-8}$ pb for $\lspone \simeq 600$ GeV whereas the LUX upper bound on $\sigma_{SI}$ is smaller by a factor of 15 for this LSP mass. However, the theoretical prediction involves several uncertainties (see Section \ref{constraint}) including the critical one - an uncertain value of $\rho_{E}$. The possibility that the predicted values are significantly weaker are, therefore , wide open. We, therefore , do not wish to exclude any model at this stage on the basis of the direct detection data.

\label{complhhs}

\subsection{LHHS Model}

\begin{figure}[h]
\centering
\includegraphics[width=0.5\textwidth]{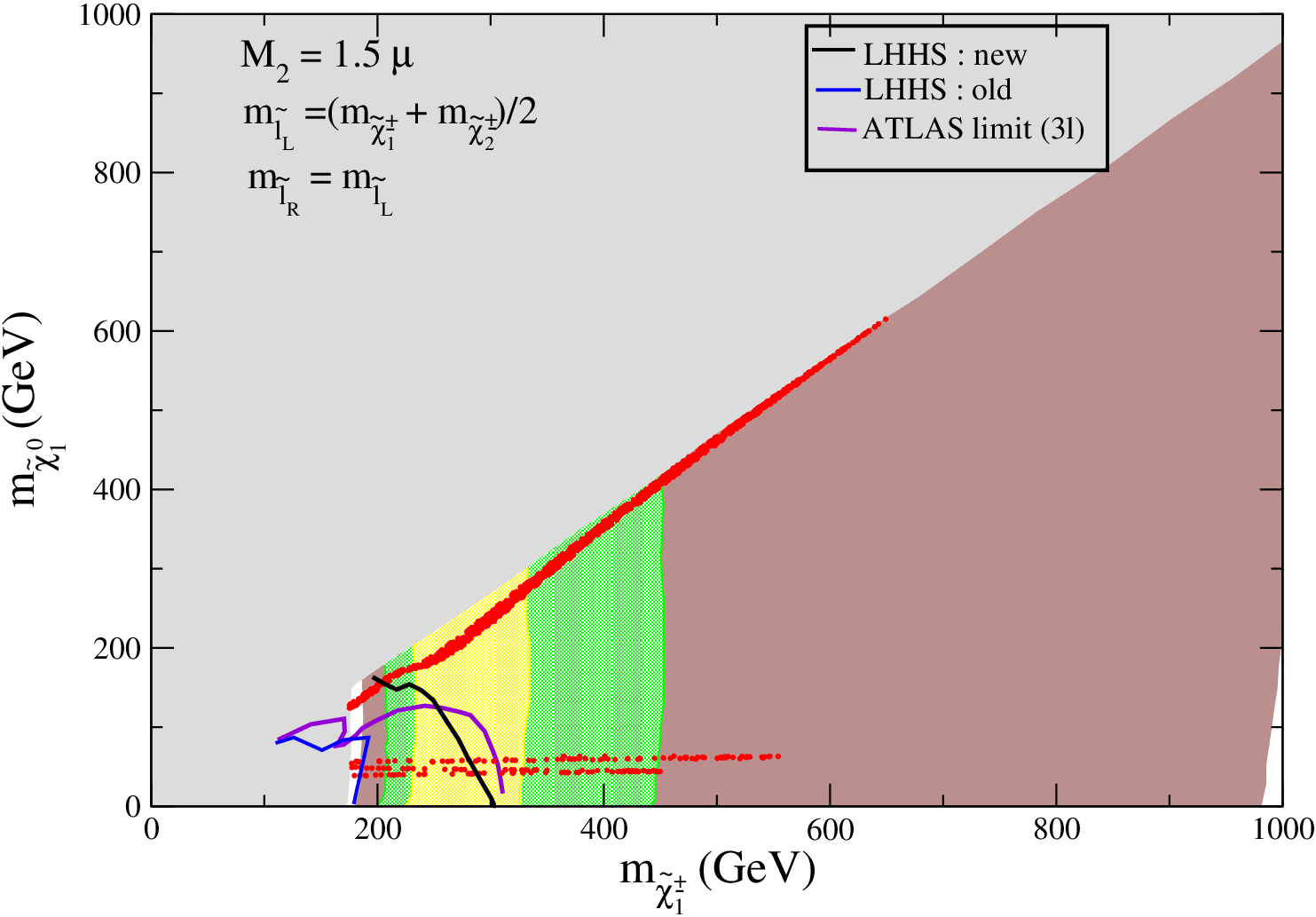}

\caption{Plot in $\mchonepm-\mlspone$ plane in Light Higgsino and Heavy Slepton (LHHS) model. The black contour is for the non-decoupled scenario whereas the blue one is for the decoupled scenario (see text for details). The magenta line is the exclusion contour as obtained by ATLAS in case of LWHS Model. Colors and conventions are same as in Fig. \ref{fig1}.}
\label{fig3}
\end{figure}

In Fig. \ref{fig3} we show our result in LHHS model (see Section \ref{lhhs}). The black line represents the exclusion contour in the non-decoupled scenario at LHC RUN I whereas the blue line represents a much weaker exclusion contour in the decoupled scenario \cite{mc2}. The magenta line represents the ATLAS exclusion in the LWHS model - the strongest limit derived in the $\mlspone-\mchonepm$ plane from RUN I data. As can be seen from the plot, the constraints are significantly  stronger due to the presence of heavier EWeakinos. For negligible LSP mass, the bound on $\mchonepm$ is found to be $\gsim 300$ GeV (earlier it was $\gsim 175$ GeV \cite{mc2}). On the other hand, for LSP mass $\gsim 165$ GeV, there is no bound on $\mchonepm$. In the decoupled scenario, the corresponding result was $\mchonepm \geq 100$ GeV. The dominant contribution to $a_{\mu}$ comes from chargino - sneutrino loop. \\

There are two separate branches consistent with the WMAP/PLANCK constraints. The dominant contributions to DM relic density in the upper branch comes from LSP pair annihilation into $W^+ W^-$ mediated by $\chonepm$, annihilation into $Z Z$ and $t \bar{t}$ through virtual $Z$ exchanges. There is also some contributions from LSP - $\chonepm$ coannihilation. In the lower branch, the main mechanism is LSP annihilation mediated by $Z$ and $h$ resonances.\\

The recent stringent constraints provided by the LUX experiment(\cite{luxrcnt}), however, question the viability of this model. The uncertainties in the prediction of $\sigma_{SI}$ limits extracted from the data as discussed in the last section \ref{constraint} does not allow a definite conclusion. For LSP masses of a few hundred GeV the upper limit on $\sigma^{SI}$ is typically of the order of a few $\times10^{-1}$ zb (i.e. $10^{-10}$ pb). On the other hand a glance at Fig. 8 of \cite{mc2} indicate that in this model the prediction is $\sim$ 10 zb in most cases. However, for LSP masses corresponding to the Z or H mediated resonant production of DM relic density, $\sigma^{SI}$ is much lower and is still consistent with the stringent LUX data.

\subsection{LHLS Model}

\begin{figure}[h]
\centering
\includegraphics[width=0.5\textwidth]{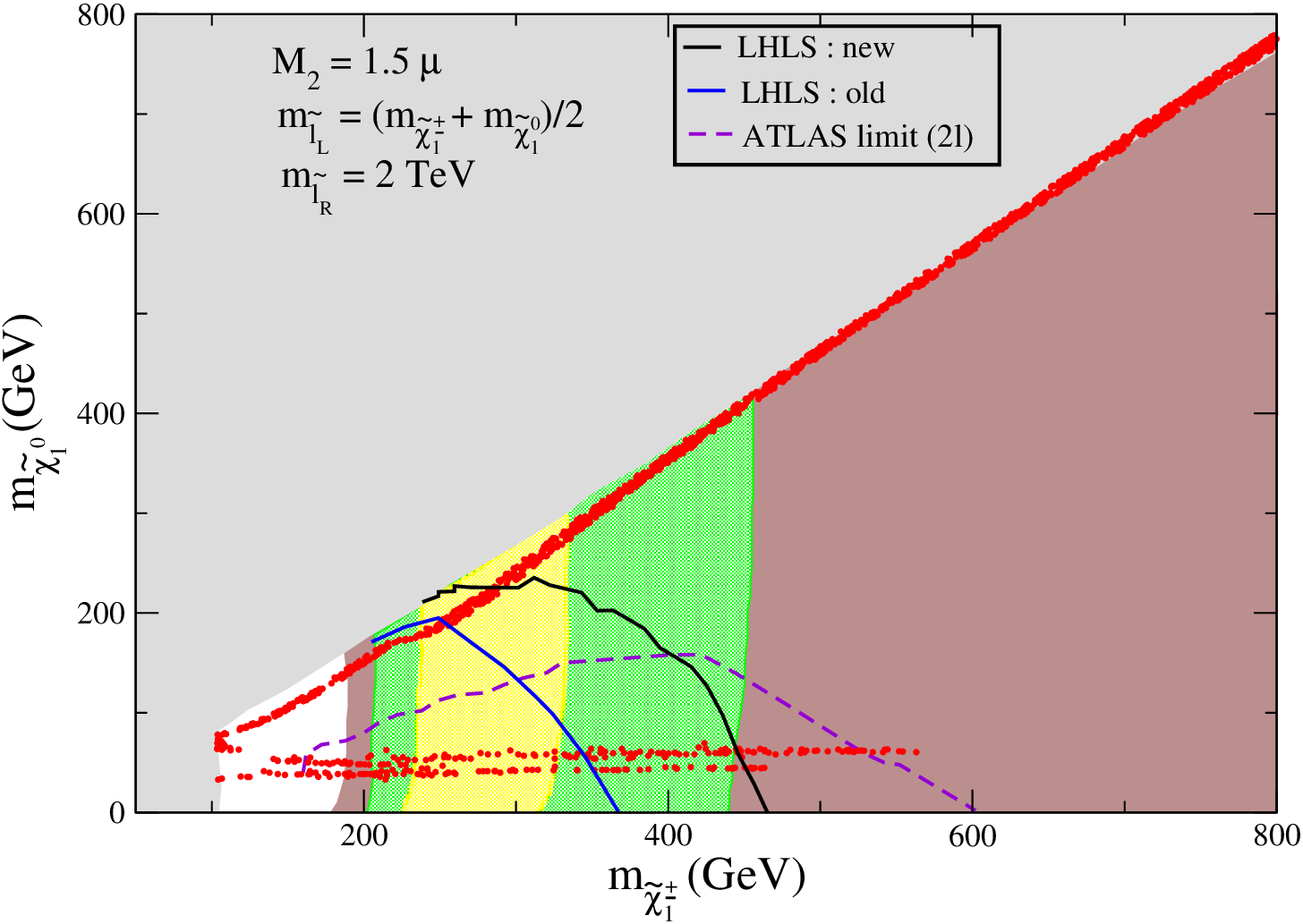}

\caption{Plot in $\mchonepm-\mlspone$ plane in Light Higgsino and light Left Slepton (LHLS) model. The black contour is for the non-decoupled scenario whereas the blue one is for the decoupled scenario (see text for details). Purple dashed line represents the exclusion contour from ATLAS slepton searches. Colors and conventions are same as in Fig. \ref{fig1}.}
\label{fig4}
\end{figure}

Fig. \ref{fig4} represents our result in LHLS model (see Section \ref{lhls}). The colour conventions for the contours are same as those in  previous section. The exclusion contour in the non-decoupled scenario is significantly stronger than the corresponding decoupled scenario. If the LSP mass is negligible, $\chonepm$ masses approximately upto $460$ GeV are excluded from LHC trilepton search. Note that, in decoupled scenario this limit was considerably weaker $\sim 365$ GeV. Also, for LSP mass $\gsim 230$ GeV (which was $\gsim 200$ GeV in the decoupled scenario), there is no bound on $\chonepm$ mass. As in the previous case , chargino - sneutrino loop dominates in case of $a_{\mu}$.  \\

$\chonepm$ mediated LSP pair annihilation into $W^+ W^-$ and annihilation into $t \bar{t}$ through virtual $Z$ exchange are the main contributing processes in the upper branch consistent with WMAP/PLANCK constraint. Small amount of annihilations into $Z Z$ and $Z h$ are also present. As $\mchonepm$ increases, $W^+ W^-$ and $t \bar{t}$ productions become subdominant and $\chonepm$ coannihilation takes over. Small amount of $\lsptwo$ coannihilation is also present. A large part of this upper branch at high $\mchonepm$ is disfavoured by the $a_{\mu}$ constraint. In the lower branch, $Z$ and $h$ production processes give the DM relic density in the right ballpark. The lower branch is strongly disfavoured by the  LHC data or by the $a_{\mu}$ constraint. From Fig. 7a of \cite{mc2} it follows that $\sigma^{SI}$ corresponding to the APS of this model violates the recent LUX bound by factors of 7-8.

\subsection{LMLS Model}
 
\begin{figure}[H]
\centering
\includegraphics[width=0.5\textwidth]{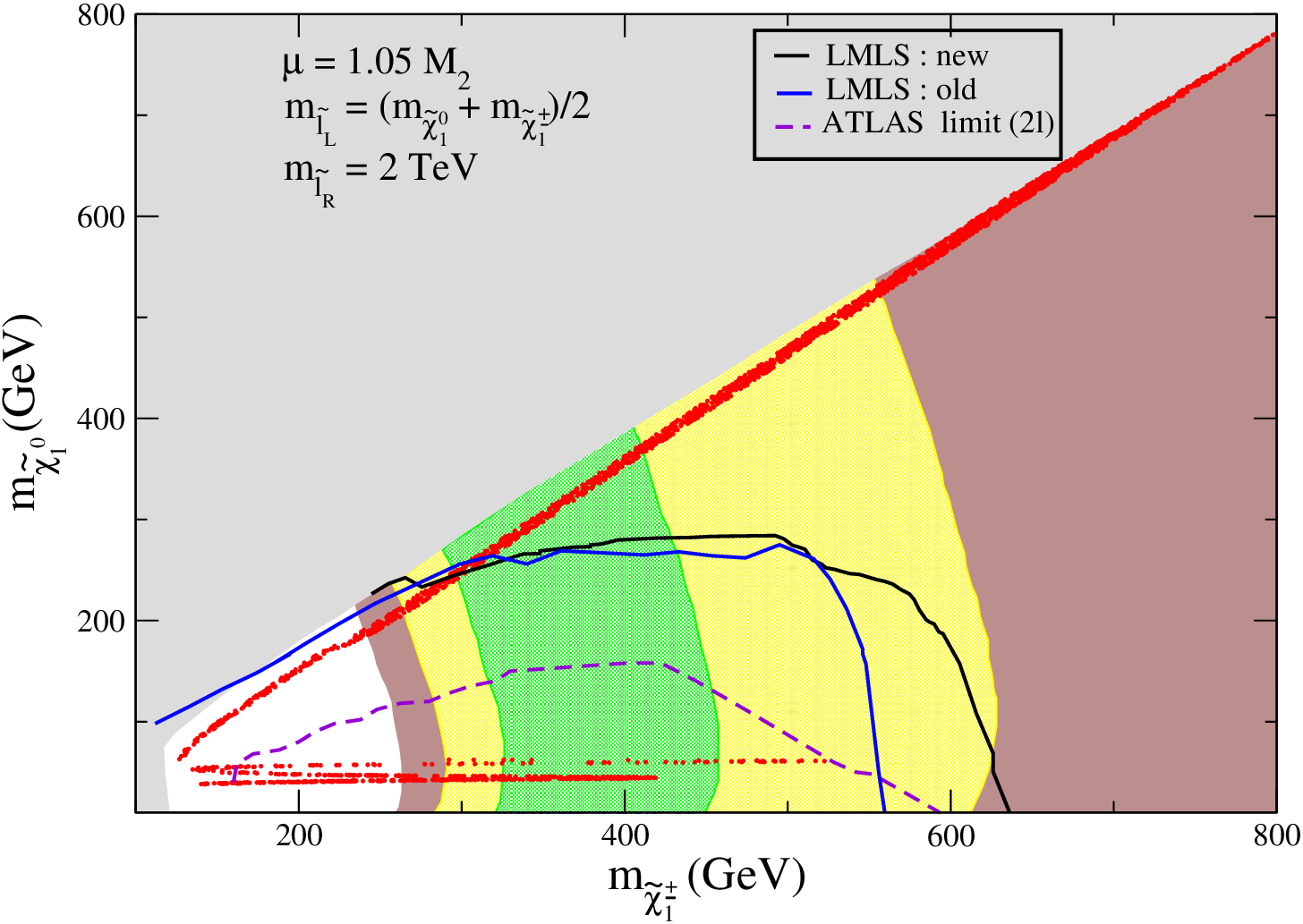}

\caption{Plot in $\mchonepm-\mlspone$ plane in Light Mixed and light Left Slepton (LMLS) model. The black contour is for the non-decoupled scenario whereas the blue one is for the decoupled scenario (see text for details). Magenta dashed line represents the exclusion contour from ATLAS slepton searches. Colors and conventions are same as in Fig. \ref{fig1}.}
\label{fig5}
\end{figure}

The APS for the LMLS model (see Section \ref{lmls}) is shown in Fig. \ref{fig5}. $\chonepm$ masses approximately upto $630$ GeV are excluded for a massless LSP whereas LHC slepton searches put a bound $\sim 600$ GeV on $\mchonepm$ for vanishing small LSP mass. The exclusion limit is considerably stronger than the decoupled scenario. Major contribution to $a_{\mu}$ is provided by chargino - sneutrino loop. \\

The WMAP/PLANCK data are satisfied by the points in the upper branch. Here DM production is mainly contributed by annihilations into $W^+ W^-$ and $t \bar{t}$ pairs. Small amount of annihilation into $Z Z$ is also present. Also non-negligible contributions come from LSP - sneutrino and LSP - slepton coannihilations. In the lower branch, $Z$ and $h$ resonance produce the correct DM relic density. In this case also, the lower branch is completely ruled out by the LHC data. A small part of the total parameter space is available which is consistent with all the constraints. The conflict between $\sigma^{SI}$ in the APS and the LUX bound persists in this model.

\section{The multilepton signatures}

In this section, we focus on the prospect of discovering several multilepton signatures in three models discussed in Section \ref{models}. As we shall show below it may be possible to discriminate against the models by the relative rates of signals in different channels. We present our results for an integrated luminosity of $100 fb^{-1}$ which is expected to accumulate before the next long shutdown of the LHC. We do not consider the LMLS model since in this case the $ml + \met$ signatures for $m > 3$ do not look very promising for the above integrated luminosity. We consider 1) $3l$, 2) $4l$, 3) SS3OS1l (three same sign and one opposite sign lepton) and 4) $5l$  final states all accompanied by large $\met$ coming from all possible EWeakino pairs - the lighter as well as the heavier ones. It may be noted that the last two signals were first studied in \cite{ng}. Here we evaluate the discovery potential of the signals after taking into account constraints derived in the last section. The observation that the heavier EWeakinos are crucial for observing final states with more than 3 leptons (see Sec \ref{hvrewkno}) will be confirmed by generator level simulation of the signals and the corresponding backgrounds for selected BPs. In our analysis we make the simplistic assumption that $S/\sqrt{B} \gsim 5$ is sufficient to claim a discovery. In case the background is negligible 5 signal events are taken as the criterion for discovery. Some of the leptonic channels have been extensively studied during RUN I of the LHC in models with decoupled heavier EWeakinos \cite{atlas3l,cms3l,atlas4l}. In contrast our emphasis in this paper is on the non-decoupled scenarios following \cite{ng}. 

For each  model, we have divided the BPs chosen for studying the multilepton signals into two categories : \\

\begin{itemize}
\item SET - I : BPs satisfying only the $3l + \met$ and $a_{\mu}$ constraints.
\item SET - II : BPs also satisfying the  WMAP/PLANCK constraints.
\end{itemize}

SET - I takes into account the possibility that there may be a non - SUSY explanation of the observed DM relic density. The BPs are enlisted in Tables \ref{tab4} and Table \ref{tab5}. All BPs are consistent with the new bounds derived in the last section.

\begin{table}[H]
\begin{center}
\begin{tabular}{|c|c|c|c|c||c|c|c|c|}
\hline
\hline
Parameters/ &\multicolumn{4}{c|}{Compressed} &\multicolumn{4}{c|}{LHHS}  \\
\cline{2-9}
Masses  &BP1 &BP2 &BP3 &BP4 &BP1 &BP2 &BP3 &BP4 \\
\cline{1-9}
$M_1$   &213  &185 &144 &134 &176 &165 &74 &32   \\
\hline
$\mu$   &223.6  &194.2 &145.9 &140.7 &240 &285 &317 &381  \\
\hline
$M_2$ &360  &466 &594 &667  &360 &427.5 &475.5  &571.5   \\
\hline
$m_{\lspone}$ &180  &155 &113   &106  &162 &157 &71 &31  \\
\hline
$m_{\chonepm}$ &213   &191 &150 &142 &228 &275 &308 &375  \\
\hline 
$m_{\chtwopm}$ &401  &500 &628 &700 &403 &469 &516 &612   \\
\hline
\hline

\end{tabular}
\end{center}
\caption{BPs consistent with the LHC and $a_{\mu}$ constraints as derived in Section \ref{newbound}. All masses and mass parameters are in GeV.}
\label{tab4}
\end{table}

\begin{table}[H]
\begin{center}
\begin{tabular}{|c|c|c|c|}
\hline
\hline
Parameters/ &\multicolumn{3}{c|}{LHLS} \\
\cline{2-4}
Masses  &BP1 &BP2 &BP3 \\
\cline{1-4}
$M_1$   &275 &241 &204 \\
\hline
$\mu$   &290  &347 &405 \\
\hline
$M_2$ &435  &520.5 &607.5 \\
\hline
$m_{\lspone}$ &244  &230 &198 \\
\hline
$m_{\chonepm}$ &280 &340 &400 \\
\hline 
$m_{\chtwopm}$ &475 &560 &646 \\
\hline
\hline

\end{tabular}
\end{center}
\caption{BPs consistent with the LHC and $a_{\mu}$ constraints as derived in Section \ref{newbound}. All masses and mass parameters are in GeV.}
\label{tab5}
\end{table}

\begin{table}[H]
\begin{center}
\begin{tabular}{|c|c|c|c|c||c|c|c|c|}
\hline
\hline
Parameters/ &\multicolumn{4}{c|}{Compressed} &\multicolumn{4}{c|}{LHHS}  \\
\cline{2-9}
Masses  &BP1-DM &BP2-DM &BP3-DM &BP4-DM &BP1-DM &BP2-DM &BP3-DM &BP4-DM \\
\cline{1-9}
$M_1$   &380 &382 &116 &116 &200 &276 &70 &60  \\
\hline
$\mu$   &399 &401.1 &121.8 &121.8 &266 &325 &349 &405 \\
\hline
$M_2$ &609  &662 &666 &736 &399 &487.5 &523.5 &607.5 \\
\hline
$m_{\lspone}$ &350  &356 &87 &88 &186 &258 &70 &60 \\
\hline
$m_{\chonepm}$ &393 &400 &123 &123 &255 &317 &342 &400 \\
\hline 
$m_{\chtwopm}$ &650  &700 &700 &770 &440 &528 &564 &647 \\
\hline
\hline

\end{tabular}
\end{center}
\caption{BPs consistent with the LHC, $a_{\mu}$ and WMAP/PLANCK constraints as derived in Section \ref{newbound}. All masses and mass parameters are in GeV.}
\label{tab6}
\end{table}

In Table \ref{tab6} the BPs are taken from both the  bands satisfying the DM relic density constraints (see Figs. \ref{fig1} and \ref{fig3}). In the LHLS model, the lower band is excluded  by the LHC and/or the $a_{\mu}$ constraints. In Table \ref{tab7}, we choose points only from the upper band.

\begin{table}[H]
\begin{center}
\begin{tabular}{|c|c|c|}
\hline
\hline
Parameters/ &\multicolumn{2}{c|}{LHLS}  \\
\cline{2-3}
Masses  &BP1-DM &BP2-DM \\
\cline{1-3}
$M_1$   &277 &403 \\
\hline
$\mu$   &328 &425 \\
\hline
$M_2$ &492 &637.5 \\
\hline
$m_{\lspone}$ &258 &376 \\
\hline
$m_{\chonepm}$ &320 &420 \\
\hline 
$m_{\chtwopm}$ &530 &676 \\
\hline
\hline

\end{tabular}
\end{center}
\caption{BPs consistent with the LHC, $a_{\mu}$ and DM relic density constraints as derived in Section \ref{newbound}. All masses and mass parameters are in GeV.}
\label{tab7}
\end{table}

\label{signal}

\subsection{The prospective $3l + \met$ signal before the next long shut down of the LHC}

The dominant SM backgrounds in this case are :

\begin{itemize}

\item $WZ$ production followed by leptonic decays of both $W$ and $Z$.
\item $ZZ$ production with $Z$ decaying into leptons where one lepton goes missing.
\item $t \bar{t}Z$ production followed by $Z \rightarrow l^+ l^-$, $t(\bar{t}) \rightarrow b(\bar{b}) W$ and one of the $W$ bosons decays into leptons.
\item $VVV$ where $(V = W,Z)$ production where leptonic decays of $W$ and $Z$ lead to trileptonic final states.

\end{itemize}

 The following sets of cuts have been used in our analysis to suppress the backgrounds :

\begin{itemize}

\item A1) Events with exactly 3 isolated leptons passing the selection cuts mentioned in section \ref{simulation} are required.
\item A2) Events with invariant mass of any Same Flavour Opposite Sign (SFOS) lepton pair falling within the window $81.2$ GeV $< m_{inv} < 101.2$ GeV are vetoed out.
\item A3) Events are required to have atleast $\met > 200$ GeV.
\item A4) Finally b-veto \cite{bveto} is applied to reduce the potentially strong background from $t \bar{t}Z$.

\end{itemize}

\begin{table}[H]
\begin{center}
\begin{tabular}{|c|c|c|c|c|c|}
\hline
\hline

Background &$\sigma_{prod}$ &\multicolumn{4}{c|}{$\sigma_{eff}^{3l}$ in $fb$} \\
\cline{3-6} 
Processes &in $pb$ &after &after &after &after\\
 & &$A1$ &$A2$ &$A3$ &$A4$\\
\cline{1-6}

$WZ$ &32.69 &168.3 &13.11 &0.18  &0.17\\
$ZZ$ &10.63 &16.5 &1.25 &0.007 &0.006\\
$t \bar{t}Z$ &0.018 &1.95 &0.39 &0.015 &0.010\\
$WWZ$ &0.133 &1.33 &0.17 &0.013 &0.012 \\
$WZZ$ &0.042 &0.54 &0.044 &0.005 &0.004 \\
$ZZZ$ &0.010 &0.05 &0.003 &0.0001 &0.0001 \\
$WWW$ &0.159 &0.79 &0.07 &0.059 &0.059 \\
\hline 
Total & & & & &\\
background &43.68 & & & &0.261\\
\hline
\hline

\end{tabular}
\end{center}
\caption{The production and effective cross-sections ($\sigma_{eff}^{3l}$) after all cuts of different SM backgrounds.}
\label{tab8}
\end{table}

\begin{table}[H]
\begin{center}
\begin{tabular}{|c|c|c|c|c|c|c|c|c|}
\hline
\hline

 &Benchmark &$\sigma_{prod}$ &\multicolumn{4}{c|}{$\sigma_{eff}^{3l}$ in $fb$} &  & \\
\cline{4-7} 
Models &Points &in $fb$ &after &after &after &after &Total $3l$ &$S/\sqrt{B}$\\
 & & &$A1$ &$A2$ &$A3$ &$A4$ &events & \\
\cline{1-9}

  &BP1 &507.9 &7.04 &5.89 &0.52 &0.48 &48.2 &9.1\\
  &BP2 &650.2 &4.28 &3.82 &0.58 &0.54 &54.6 &10.3\\
  &BP3 &1417. &4.84 &4.59 &0.60 &0.58 &58.1 &10.9 \\
  &BP4 &1763. &4.05 &4.01 &0.27 &0.26 &26.4 &4.9\\
\cline{2-9}
Compressed   &BP1-DM &56.38 &0.67 &0.57 &0.13 &0.12 &12.1 &2.3\\
  &BP2-DM &51.21 &0.49 &0.42 &0.12 &0.10 &10.8 &2.0\\
  &BP3-DM &2817. &5.52 &5.27 &0.34 &0.33 &33.8 &6.4\\
  &BP4-DM &2812. &6.10 &6.04 &0.20 &0.16 &16.8 &3.2\\
\hline
  &BP1 &555.4 &6.80 &0.78 &0.67 &0.66 &66.6 &12.6\\
  &BP2 &301.4 &4.53 &0.74 &0.49 &0.49 &49.1 &9.3 \\
  &BP3 &200.5 &2.62 &0.97 &0.57 &0.57 &57.5 &10.8\\
  &BP4 &92.56 &1.16 &0.59 &0.30 &0.30 &30.8 &5.8\\
\cline{2-9}
LHHS   &BP1-DM &370.8 &4.34 &0.48 &0.35 &0.35 &35.2 &6.7 \\
  &BP2-DM &156.3 &1.84 &0.29 &0.22 &0.22 &22.0 &4.2 \\
  &BP3-DM &134.8 &1.64 &0.79 &0.46 &0.46 &46.9 &8.9 \\
  &BP4-DM &71.12 &0.84 &0.45 &0.25 &0.25 &25.4 &4.8 \\
\hline
  
\hline
\hline

\end{tabular}
\end{center}
\caption{The production cross-sections of all EWeakino pairs and $\sigma_{eff}^{3l}$ for the BPs defined in Table \ref{tab4} and Table \ref{tab6}. Also the total number of $3l$ events along with $S/\sqrt{B}$ are shown for an integrated luminosity of $100 fb^{-1}$.}
\label{tab9}
\end{table}

We present the estimated number of background and signal events for an integrated luminosity of $100 fb^{-1}$ in Tables \ref{tab8} - \ref{tab10}. For the compressed model, all the BPs corresponding to SET - I can lead to discovery for the quoted luminosity. However, for SET - II  the $3l$ signal is rather weak except for BP3-DM. The others may be relevant as higher luminosities accumulate after the next long shutdown. For LHHS and LHLS Model, both sets of BPs (SET - I and SET - II) give observable $3l$ signal.


\begin{table}[H]
\begin{center}
\begin{tabular}{|c|c|c|c|c|c|c|c|c|}
\hline
\hline

 &Benchmark &$\sigma_{prod}$ &\multicolumn{4}{c|}{$\sigma_{eff}^{3l}$ in $fb$} &  & \\
\cline{4-7} 
Models &Points &in $fb$ &after &after &after &after &Total $3l$ &$S/\sqrt{B}$\\
 & & &$A1$ &$A2$ &$A3$ &$A4$ &events & \\
\cline{1-9}
  &BP1 &210.1 &9.9 &9.1 &0.96 &0.92 &92.0 &17.4 \\
  &BP2 &131.5 &8.66 &5.97 &1.02 &0.98 &98.5 &18.6 \\
  &BP3 &69.93 &3.02 &2.41 &0.82 &0.78 &78.3 &14.8 \\
\cline{2-9}
LHLS   &BP1-DM &151.8 &111.2 &102.2 &1.04 &0.98 &98.5 &18.6 \\
  &BP2-DM &45.25 &2.26 &2.09 &0.35 &0.32 &32.5 &6.15 \\
\hline
\hline

\end{tabular}
\end{center}
\caption{The production cross-sections of all EWeakino pairs and $\sigma_{eff}^{3l}$ for the BPs defined in Table \ref{tab5} and Table \ref{tab7}. Also the total number of $3l$ events along with $S/\sqrt{B}$ are shown for an integrated luminosity of $100 fb^{-1}$.}
\label{tab10}
\end{table}

\label{trilep}

\subsection{The prospective $4l + \met$ signal}
In this section we assess the discovery potential of the $4l$ channel at LHC RUN II. It may be noted that our analysis based on the pMSSM is more general than the ATLAS 4l analysis \cite{atlas4l} in a simplified model. The differences between the two approaches have been discussed in \cite{ng}. The heavier EWeakinos play pivotal role in this case (see Section \ref{hvrewkno}). The main SM backgrounds are :

\begin{itemize}

\item $ZZ$ production where both $Z$ decay leptonically.
\item $t \bar{t}Z$ production followed by $Z \rightarrow l^+ l^-$ and leptonic decays of the $W^{\pm}$ bosons coming from the top decay.
\item $VVV$ with $Z$ and $W^{\pm}$ decaying into leptons.

\end{itemize} 

The size of the SM background is considerably smaller than that for the trilepton final states. We apply the following set of cuts to make the background negligible :

\begin{itemize}

\item B1) Exactly 4 isolated leptons passing all the selection cuts (see Section \ref{simulation}) are required.
\item B2) The invariant mass of any SFOS lepton pair should not fall within the window $81.2 - 101.2$ GeV.
\item B3) Events must have $\met > 80$ GeV.
\item B4) A b-veto is applied to suppress the background coming from $t \bar{t}Z$.

\end{itemize}

\begin{table}[H]
\begin{center}
\begin{tabular}{|c|c|c|c|c|c|}
\hline
\hline

Background &$\sigma_{prod}$ &\multicolumn{4}{c|}{$\sigma_{eff}^{4l}$ in $fb$} \\
\cline{3-6} 
Processes &in $pb$ &after &after &after &after\\
 & &$B1$ &$B2$ &$B3$ &$B4$\\
\cline{1-6}

$ZZ$ &10.63 &14.2 &0.081 &0 &0\\
$t \bar{t}Z$ &0.018 &0.26 &0.039 &0.018 &0.005 \\
$WWZ$ &0.133 &0.18 &0.012 &0.004 &0.002 \\
$WZZ$ &0.042 &0.068 &0.0014 &0.0003 &0.0003 \\
$ZZZ$ &0.010 &0.04 &0.0003 &0.00005 &0.00005 \\
\hline 
\hline

\end{tabular}
\end{center}
\caption{The production and effective cross-sections ($\sigma_{eff}^{4l}$) after the cuts for different SM backgrounds.}
\label{tab11}
\end{table}

\begin{table}[H]
\begin{center}
\begin{tabular}{|c|c|c|c|c|c|c|c|}
\hline
\hline

 &Benchmark &$\sigma_{prod}$ &\multicolumn{4}{c|}{$\sigma_{eff}^{4l}$ in $fb$} &   \\
\cline{4-7} 
Models &Points &in $fb$ &after &after &after &after &Total $4l$ \\
 & & &$B1$ &$B2$ &$B3$ &$B4$ &events  \\
\cline{1-8}

  &BP1 &507.9 &1.18 &0.78 &0.48 &0.46 &46.7 \\
  &BP2 &650.2 &0.56 &0.42 &0.32 &0.29 &29.2 \\
  &BP3 &1417. &0.11 &0.07 &0.07 &0.07 &7.08 \\
  &BP4 &1763. &0.09 &0.07 &0.03 &0.03 &3.52 \\
\cline{2-8}
Compressed   &BP1-DM &56.38 &0.10 &0.06 &0.05 &0.04 &4.62 \\
  &BP2-DM &51.21 &0.05 &0.03 &0.02 &0.02 &2.61 \\
  &BP3-DM &2817. &0.22 &0.19 &0.14 &0.14 &14.1 \\
  &BP4-DM &2812. &0.14 &0.11 &0.05 &0.05 &5.62 \\
\hline
  &BP1 &555.4 &0.70 &0.45 &0.25 &0.24 &24.9 \\
  &BP2 &301.4 &0.41 &0.32 &0.10 &0.10 &10.2 \\
  &BP3 &200.5 &0.22 &0.18 &0.06 &0.06 &6.41 \\
  &BP4 &92.56 &0.08 &0.07 &0.03 &0.02 &2.86 \\
\cline{2-8}
LHHS   &BP1-DM &370.8 &0.52 &0.33 &0.18 &0.18 &18.5  \\
  &BP2-DM &156.3 &0.19 &0.15 &0.10 &0.10 &10.1  \\
  &BP3-DM &134.8 &0.13 &0.12 &0.04 &0.03 &3.64  \\
  &BP4-DM &71.12 &0.07 &0.07 &0.02 &0.02 &2.34  \\
\hline
\hline
  
\hline

\end{tabular}
\end{center}
\caption{The production cross-sections of all EWeakino pairs and $\sigma_{eff}^{4l}$ for the BPs defined in Table \ref{tab4} and Table \ref{tab6}. Also the total number of $4l$ events are shown for an integrated luminosity of $100 fb^{-1}$.}
\label{tab12}
\end{table}

\begin{table}[H]
\begin{center}
\begin{tabular}{|c|c|c|c|c|c|c|c|}
\hline
\hline

 &Benchmark &$\sigma_{prod}$ &\multicolumn{4}{c|}{$\sigma_{eff}^{4l}$ in $fb$} &   \\
\cline{4-7} 
Models &Points &in $fb$ &after &after &after &after &Total $4l$ \\
 & & &$B1$ &$B2$ &$B3$ &$B4$ &events  \\
\cline{1-8}

  &BP1 &210.1 &0.26 &0.19 &0.12 &0.11 &11.7 \\
  &BP2 &131.5 &0.44 &0.17 &0.11 &0.10 &10.9  \\
  &BP3 &69.93 &0.12 &0.05 &0.04 &0.04 &4.40 \\
\cline{2-8}
LHLS   &BP1-DM &151.8 &0.39 &0.29 &0.15 &0.13 &13.8 \\
  &BP2-DM &45.25 &0.09 &0.06 &0.04 &0.03 &3.8 \\
\hline
\hline
 
\end{tabular}
\end{center}
\caption{The production cross-sections of all EWeakino pairs and $\sigma_{eff}^{4l}$ for the BPs  defined in Table \ref{tab5} and Table \ref{tab7}. Also the total number of $4l$ events are shown for an integrated luminosity of $100 fb^{-1}$.}
\label{tab13}
\end{table} 

Tables \ref{tab11} - \ref{tab13} summarize the results. The total background is found to be vanishingly small after the cuts. For the compressed model, most of the BPs belonging to both the sets indicate potential discovery chances for $L = 100 fb^{-1}$. In case of LHHS model, BPs of SET - I can give rise to large $4l + \met$ signal except the last one. For WMAP/PLANCK data satisfying points, the result is weaker for comparatively heavy $\chonepm$. Finally for LHLS model, the BPs of both sets lead to sufficiently large $4l + \met$ signal. 

\label{fourlep}

\subsection{Three Same Sign and One Opposite Sign Leptons (SS3OS1) $+ \met$ signal}

We now discuss a special case of $4l + \met$ signal when the total charge of the final state leptons is necessarily non-zero. This is of particular interest as the corresponding SM background is very small and can be made to vanish by applying moderate cuts. Other interesting features will be discussed below. The main backgrounds are :

\begin{itemize}

\item $t \bar{t} Z$ production.
\item $WZZ$ production followed by leptonic decays of all the gauge bosons where one lepton fails to pass the selection cuts.
\item $ZZZ$ production.

\end{itemize}

Requiring 4 isolated leptons with non-zero total charge (C1)) and $\met > 80$ GeV (C2)) are found to be effective for reducing the background significantly. This is shown in Table \ref{tab14}.

\begin{table}[H]
\begin{center}
\begin{tabular}{|c|c|c|c|}
\hline
\hline

Background &$\sigma_{prod}$ &\multicolumn{2}{c|}{$\sigma_{eff}^{ss3os1}$ in $fb$} \\
\cline{3-4} 
Processes &in $pb$ &after &after \\
 & &$C1$ &$C2$ \\
\cline{1-4}

$t \bar{t}Z$ &0.018 &0.006 &0.002   \\
$WZZ$ &0.042 &0.007 &0.003  \\
$ZZZ$ &0.010 &0.0004 &0.00003 \\
\hline 
\hline

\end{tabular}
\end{center}
\caption{The production and effective cross-sections ($\sigma_{eff}^{ss3os1}$) after the cuts  of different SM backgrounds.}
\label{tab14}
\end{table}  

In Table \ref{tab15} - \ref{tab16}, we show the number of signal events surviving the successive cuts.

\begin{table}[H]
\begin{center}
\begin{tabular}{|c|c|c|c|c|c|}
\hline
\hline

 &Benchmark &$\sigma_{prod}$ &\multicolumn{2}{c|}{$\sigma_{eff}^{ss3os1}$ in $fb$} &   \\
\cline{4-5} 
Models &Points &in $fb$ &after &after &Total SS3OS1l \\
 & & &$D1$ &$D2$ &events  \\
\cline{1-6}

  &BP1 &507.9 &0.29 &0.20 &20.3  \\
  &BP2 &650.2 &0.17 &0.15 &15.6  \\
  &BP3 &1417. &0.014 &0.014 &1.42  \\
  &BP4 &1763. &0.035 &0.017 &1.76  \\
\cline{2-6}
Compressed   &BP1-DM &56.38 &0.033 &0.025 &2.53  \\
  &BP3-DM &2817. &0.084 &0.056 &5.63  \\
\hline
  &BP1 &555.4 &0.11 &0.083 &8.33  \\
  &BP2 &301.4 &0.045 &0.039 &3.92  \\
  &BP3 &200.5 &0.028 &0.020 &2.01  \\
  &BP4 &92.56 &0.017 &0.013 &1.38  \\
\cline{2-6}
LHHS   &BP1-DM &370.8 &0.056 &0.033 &3.34  \\
  &BP2-DM &156.3 &0.036 &0.028 &2.81   \\
  &BP3-DM &134.8 &0.017 &0.016 &1.62   \\
  &BP4-DM &71.12 &0.011 &0.010 &1.06   \\
\hline
  
\hline

\end{tabular}
\end{center}
\caption{The production cross sections of all EWeakino pairs and $\sigma_{eff}^{ss3os1}$ after successive cuts for the BPs defined in Table \ref{tab5} and Table \ref{tab7}. Also the total number of SS3OS1l events are shown for an integrated luminosity of $100 fb^{-1}$.}
\label{tab15}
\end{table}

\begin{table}[h]
\begin{center}
\begin{tabular}{|c|c|c|c|c|c|}
\hline
\hline

 &Benchmark &$\sigma_{prod}$ &\multicolumn{2}{c|}{$\sigma_{eff}^{ss3os1}$ in $fb$} &   \\
\cline{4-5} 
Models &Points &in $fb$ &after &after &Total SS3OS1l \\
 & & &$D1$ &$D2$ &events  \\
\cline{1-6}

  &BP1 &210.1 &0.05 &0.04 &3.99  \\
  &BP2 &131.5 &0.03 &0.028 &2.36 \\
\cline{2-6}
LHLS   &BP1-DM &151.8 &0.04 &0.03 &2.88   \\
  &BP2-DM  &45.25 &0.013 &0.012 &1.18 \\
\hline
\hline

\end{tabular}
\end{center}
\caption{The production cross sections of all EWeakino pairs and $\sigma_{eff}^{ss3os1}$ after successive cuts for the BPs  defined in Table \ref{tab6} and Table \ref{tab8}. Also the total number of SS3OS1l events are shown for an integrated luminosity of $100 fb^{-1}$.}
\label{tab16}
\end{table}

The background is practically zero after one applies the above cuts. Tables \ref{tab13} - \ref{tab15} show our results for SS3OS1l signals. For compressed model, it is possible to get more than 5 signal events for some of the BPs in both SET - I and SET - II. Although in LHHS (except BP1) and LHLS models the SS3OS1l $+ \met$ signal events never reach 5 for the considered value of integrated luminosity, we would like to mention that some of them may as well serve as a hint. Thus the observation of this signal before the next long shut down of the LHC may reduce the number of competing models.

\label{ss3os1lep}

\subsection{$5l + \met$ signal}

Next we discuss the prospects of observing 5 isolated leptons associated with missing energy in the final state coming from EWeakino productions at RUN II of LHC. A few SM processes can give rise to the corresponding background. We enlist them below :

\begin{itemize}

\item $t \bar{t} Z$ production where both $Z$ and $W^{\pm}$ bosons (coming from top decays) decay into leptons and the remaining one comes from a heavy quark decay.
\item $WZZ$ production followed by leptonic decays of all the gauge bosons.
\item $ZZZ$ production with leptonic decay of all $Z$ bosons.

\end{itemize}

Demanding 5 isolated leptons in the final state (D1)) itself reduces the number of background events significantly. A moderate cut $\met > 80$ GeV (D2)) is then sufficient to bring it down to a negligible level. The effect of the cuts on the SM processes is illustrated in Table \ref{tab17}.

\begin{table}[H]
\begin{center}
\begin{tabular}{|c|c|c|c|}
\hline
\hline

Background &$\sigma_{prod}$ &\multicolumn{2}{c|}{$\sigma_{eff}^{5l}$ in $fb$} \\
\cline{3-4} 
Processes &in $pb$ &after &after \\
 & &$D1$ &$D2$ \\
\cline{1-4}

$t \bar{t}Z$ &0.018 &0.002 &0.0007   \\
$WZZ$ &0.042 &0.013 &0.005  \\
$ZZZ$ &0.010 &0.001 &0.0003 \\
\hline 
\hline

\end{tabular}
\end{center}
\caption{The production level and effective cross-sections ($\sigma_{eff}^{5l}$) after the cuts of different backgrounds.}
\label{tab17}
\end{table}

We quote the number of signal events in Table \ref{tab18} - \ref{tab19}.

\begin{table}[H]
\begin{center}
\begin{tabular}{|c|c|c|c|c|c|}
\hline
\hline

 &Benchmark &$\sigma_{prod}$ &\multicolumn{2}{c|}{$\sigma_{eff}^{5l}$ in $fb$} &   \\
\cline{4-5} 
Models &Points &in $fb$ &after &after &Total $5l$ \\
 & & &$D1$ &$D2$ &events  \\
\cline{1-6}

  &BP1 &507.9 &0.16 &0.096 &9.65  \\
  &BP2 &650.2 &0.06 &0.052 &5.20  \\
  &BP3 &1417. &0.03 &0.028 &2.83  \\
  &BP4 &1763. &0.02 &0.017 &1.76  \\
\cline{2-6}
Compressed   &BP1-DM &56.38 &0.014 &0.012 &1.20  \\
  &BP2-DM &2817. &0.03 &0.028 &2.81  \\
\hline
  &BP1 &555.4 &0.11 &0.077 &7.77  \\
  &BP2 &301.4 &0.05 &0.039 &3.91  \\
  &BP3 &200.5 &0.02 &0.018 &1.80  \\
  &BP4 &92.56 &0.013 &0.012 &1.20  \\
\cline{2-6}
LHHS   &BP1-DM &370.8 &0.06 &0.041 &4.10  \\
  &BP2-DM &156.3 &0.025 &0.015 &1.56   \\
  &BP3-DM &134.8 &0.011 &0.011 &1.13   \\
  &BP4-DM &71.12 &0.016 &0.014 &1.42   \\
\hline
\hline
  
\hline

\end{tabular}
\end{center}
\caption{The production cross sections of all EWeakino pairs and $\sigma_{eff}^{5l}$ for the BPs defined in Table \ref{tab4} and Table \ref{tab6}. Also the total number of $5l$ events are shown for an integrated luminosity of $100 fb^{-1}$.}
\label{tab18}
\end{table}

\begin{table}[H]
\begin{center}
\begin{tabular}{|c|c|c|c|c|c|}
\hline
\hline

 &Benchmark &$\sigma_{prod}$ &\multicolumn{2}{c|}{$\sigma_{eff}^{5l}$ in $fb$} &   \\
\cline{4-5} 
Models &Points &in $fb$ &after &after &Total $5l$ \\
 & & &$D1$ &$D2$ &events  \\
\cline{1-6}

  &BP1 &210.1 &0.039 &0.035 &3.57  \\
  &BP2 &131.5 &0.025 &0.018 &1.84  \\
  &BP3 &69.93 &0.012 &0.01 &1.04   \\
\cline{2-6}
LHLS   &BP1-DM &151.8 &0.044 &0.036 &3.64  \\
  &BP2-DM &45.25 &0.014 &0.012 &1.08  \\
\hline

\end{tabular}
\end{center}
\caption{The production cross-sections of all EWeakino pairs and $\sigma_{eff}^{5l}$ for the BPs defined in Table \ref{tab5} and Table \ref{tab7}. Also the total number of $5l$ events are shown for an integrated luminosity of $100 fb^{-1}$.}
\label{tab19}
\end{table}

Note that in this case signal events in WMAP/PLANCK allowed points never reach 5 for $100 fb^{-1}$ of integrated luminosity though some of them may show up as a hint in the early phases of RUN II. Therefore, one has to wait for upgradation in luminosity to claim a discovery through this channel.

\label{fivelep}

\subsection{Multilepton signals in moderately compressed LHHS models}

As already discussed in Section \ref{complhhs} and confirmed in Section \ref{trilep}, there is a tension between the DM relic density constraint and low $\mchtwopm$ in a highly compressed scenario characterised by the representative choice $\mu = 1.05 M_1$. Relaxing the degree of compression one obtains consistency with the observed DM relic density for much lower $\mchtwopm$ (see Table \ref{tab20}). 

\begin{table}[H]
\begin{center}
\begin{tabular}{|c|c|c|c|}
\hline
\hline

Points &$\mu = x M_1$ &$\lspone = 200$, $\chtwopm = 400$ &$\Omega_{\tilde{\chi}}$ \\
\hline
1 &$x=1.05$ &$\chonepm = 231$ &0.0238 \\
2 &$x=1.20$ &$\chonepm = 248$ &0.0566 \\
3 &$x=1.30$ &$\chonepm = 260$ &0.0975 \\
\hline
\hline
Points &$\mu = x M_1$ &$\lspone = 300$, $\chtwopm = 500$ &$\Omega_{\tilde{\chi}}$ \\
\hline
4 &$x=1.05$ &$\chonepm = 335$ &0.044 \\
5 &$x=1.10$ &$\chonepm = 342$ &0.0669 \\
6 &$x=1.15$ &$\chonepm = 350$ &0.102 \\
\hline
\hline

\end{tabular}
\end{center}
\caption{DM relic densities for two different $\chtwopm$ masses in relaxed compression scenario. All masses are in GeV.}
\label{tab20}
\end{table}

It follows from Table \ref{tab9} that the parameter space consistent with the DM relic density data may not yield a $3l + \met$ signal. On the other hand for $\mu = 1.3 M_1$ it follows from Table \ref{tab21} that encouraging $3l$ signals are predicted in all cases. Other background free multilepton signals also look promising. 

\begin{table}[H]
\begin{center}
\begin{tabular}{|c|c|c|c|c|c|c|c|}
\hline
\hline

 &\multicolumn{3}{c|}{Masses} &\multicolumn{4}{c|}{Signals} \\
\cline{2-4} \cline{5-8} 

Points &$\lspone$ &$\chonepm$ &$\chtwopm$ &$(S/\sqrt{B})_{3l}$ &4l &SS3OS1l &5l  \\
\cline{1-8}
1 &200 &265 &420 &6.2 &22.5 &9.6 &5.3  \\
2 &200 &275 &500 &7.2 &13.7 &3.5 &4.1 \\
3 &250 &300 &420 &3.7 &15.7 &4.6 &6.1 \\

\hline
\hline

\end{tabular}
\end{center}
\caption{Number of events surviving all cuts for all types of signals for an integrated luminosity of 100 $fb^{-1}$ in the moderately compressed model. Masses are given in GeV.}
\label{tab21}
\end{table} 

\label{relaxedcomp}

\subsection{Discriminating different models via multilepton signatures}

As pointed out in \cite{ng} if more than one multilepton signatures show up before  the next long shutdown of the LHC, their relative rates  may distinguish  different models studied in this paper. This can be showcased by the compressed model. For BP1-DM, BP2-DM and BP3-DM the $3l$ signal is unobservable whereas for the first and the last BP the $4l$ signal is likely to be observed. It is interesting to note that for all other BPs the 3l signal is observable. Thus if  the 4l signal (and not the 3l signal)  is observed BP1-DM and BP3-DM would be strong candidates for the underlying model. These two models, in turn, can be distinguished since only for BP3-DM  the SS3OS1l signal is observable. 

In fact the SS3OS1l signal could be a useful discriminator for the models. Both the 3l and SS3OS1l signals can be observed in the compressed model (BP1, BP2) and the LHHS model (BP1). The ratio $r_{ss/4l} = $ (the number of SS3OS1l events) / (the number of 4l events) is approximately 1/3 (1/2) for LHHS (BP1) (compressed(BP2)) model. On the other hand for compressed model (BP1) the ratio $r_{4l/3l} = $ (the number of 4l events) / (the number of 3l events) is roughly 1.4 while the same ratio is significantly smaller than 1 for compressed (BP2) and LHHS (BP1).

This procedure can be employed for distinguishing the BPs presented in Tables \ref{tab4} - \ref{tab7} from each other. Obviously the method will be more efficient as luminosity accumulates at the LHC and reduces the statistical errors. Some of the systematics like uncertainties in the production cross-sections cancel if we consider the relative rates.

\label{discriminate}

\section{Conclusion}

In order to extend and complement \cite{ng}, we have examined the complete EWeakino sector of several pMSSM medels with our main attention focussed on the heavier EWeakinos.

In Section \ref{models} we have argued that in view of the $3l$ + ${E\!\!\!\!/_T}$ data and the naturalness arguments models where the heavier EWeakinos are wino like, the lighter ones are higgsino like and the LSP is either bino like or a bino-higgsino admixture are  preferred for interesting phenomenology. Accordingly we have targeted the compressed model (Section \ref{complhhs}), the LHHS model (Section \ref{lhhs}), the LHLS model (Section \ref{lhls}) and the LMLS model (Section \ref{lmls}).

In Section \ref{hvrewkno} we have computed the production cross-sections of all EWeakinos pairs and the BRs of each EWeakino in the above models to illustrate that the multilepton ($ml$) + ${E\!\!\!\!/_T}$ final states for $ m > 3 $ are viable only if the heavier EWeakinos are not decoupled (See Table \ref{tab3}).

In Section \ref{method} we describe our methodology. The constraints that we used in our analyses are summarized in section \ref{constraint} . We do not consider some often used constraints, most notably the direct DM detection data, since they involve sizable uncertainties. However, we have shown in subsequent sections that the models studied by us are compatible with the data if allowances are made for these uncertainties.

In Section \ref{newbound} we delineated the allowed parameter space in each case subject to the constraints from the LHC $3l$ + ${E\!\!\!\!/_T}$ data, the observed DM relic density of the universe and the precisely measured anomalous magnetic moment of the muon at the $\lsim 2\sigma $ level (See Figs. \ref{fig1} - \ref{fig4}). The largest parameter space is allowed in the compressed model ($\mu=1.05 M_1$). However, if consistency with the DM relic density constraint requires that $m_{\chtwopm} \ge 600$ GeV irrespective of $m_{\lspone}$. This may adversely affect the observability of some potential multilepton signatures. If the compression is relaxed lower values of $m_{\chtwopm}$ are allowed (See below).

In Section \ref{signal} we select benchmark points (BPs) from the APS of each model delineated in Section \ref{newbound} (See Figs. \ref{fig1} - \ref{fig4}). We show that in most cases observable $3l$ (Section \ref{trilep}), $4l$ (Section \ref{fourlep}), $SS3OS1$ (Section \label{ss3os1l}) and $5l$ (Section \ref{fivelep}) signal accompanied by large ${E\!\!\!\!/_T}$ can all show up before the next long shut down of the LHC. None of the signals are viable if the heavier EWeakinos are decoupled.

We show in Section \ref{relaxedcomp} that if the compression is relaxed, smaller $m_{\chtwopm}$ ($\lsim 600$ GeV) is compatible with DM relic density constraint (See Table \ref{tab20}) and observable multilepton signals are viable (Table \ref{tab21}).  
  
In Section \ref{discriminate} we discuss the prospect of discriminating between competing models using the relative rates of different multilepton signatures in these models.

\label{conclude}
{ \bf Acknowledgments :} MC would like to thank TRR33 ”The Dark Universe” project for financial support. NG thanks Science and Engineering Research Board, Department of Science and Technology, India for a research fellowship. 


\end{document}